\documentclass{elsart}

\usepackage{amsmath, amssymb, epic, jump}
\usepackage{graphicx}

\newcommand{\ie}{\textit{i.e.}}

\journal{Theoretical Computer Science}

\begin{document}

  \newcommand{\cell}[1]{\makebox[0.25cm]{#1}}
  \newcommand{\sdot}{{\scriptstyle \ldots}\;}

  \newcommand{\CA}{$\mathcal{CA}\,$}
  \newcommand{\CAa}[2]{$\mathcal{CA}(#1,#2)$}
  \newcommand{\NCCA}{$\mathcal{CA}^0\,$}
  \newcommand{\NCCAa}[2]{$\mathcal{CA}^0(#1,#2)$}
  \newcommand{\DCA}{$\mathcal{CA}^-\,$}
  \newcommand{\DCAa}[2]{$\mathcal{CA}^-(#1,#2)$}
  \newcommand{\ICA}{$\mathcal{CA}^+\,$}
  \newcommand{\ICAa}[2]{$\mathcal{CA}^+(#1,#2)$}

  \newcommand{\PA}{$\mathcal{PA}\,$}
  \newcommand{\PAa}[2]{$\mathcal{PA}(#1,#2)$}
  \newcommand{\NCPA}{$\mathcal{PA}^0\,$}
  \newcommand{\NCPAa}[2]{$\mathcal{PA}^0(#1,#2)$}

  \newcounter{example}

\begin{frontmatter}

  \title{On Conservative and Monotone \\ 
         One-dimensional Cellular Automata and \\
         Their Particle Representation}

  \author{Andr\'es Moreira}
  \address{Center for Mathematical Modeling and Departamento 
           de Ingenier\'{\i}a Matem\'atica \\
           FCFM, U. de Chile, Casilla 170/3-Correo 3, Santiago, Chile}
  \ead{anmoreir@dim.uchile.cl}
  \author{Nino Boccara}
  \address{Department of Physics, University of Illinois, Chicago, USA\\
    and DRECAM/SPEC, CE Saclay, 91191 Gif-sur-Yvette Cedex, France}
  \author{Eric Goles}
  \address{Center for Mathematical Modeling and Departamento 
           de Ingenier\'{\i}a Matem\'atica \\
           FCFM, U. de Chile, Casilla 170/3-Correo 3, Santiago, Chile}

  \begin{abstract}
   Number-conserving (or {\em conservative}) cellular automata 
	have been used in several
    contexts, in particular traffic models, where it is natural to 
    think about
    them as systems of interacting particles. In this article
	 we consider several issues concerning one-dimensional 
	 cellular automata which are conservative, monotone (specially
	 ``non-increasing''), or that allow a weaker kind of conservative
	 dynamics. We introduce a formalism of ``particle automata'',
	 and discuss several properties that they may exhibit, some of which, like
    anticipation and momentum preservation, happen to be intrinsic to
    the conservative CA they represent. 
	 For monotone CA we give a characterization, and then show that
	 they too are equivalent to the corresponding class of particle
	 automata. Finally, we show how to determine, for a given CA
    and a given integer $b$, whether its states admit a
    $b$-neighborhood-dependent relabelling whose sum is conserved
    by the CA iteration; this can be used to uncover conservative
    principles and particle-like behavior underlying the dynamics
    of some CA.
	 \\
	 Complements at {\tt http://www.dim.uchile.cl/\verb'~'anmoreir/ncca}
  \end{abstract}

\end{frontmatter}

\section{Introduction}

  Cellular automata (CA) are discrete dynamical systems, where 
  {\em states}
  taken from a finite set of possible values are assigned to each
  cell of some regular lattice;
  at each time step, the state of a cell is updated through
  a function whose inputs are the states of the
  cell and its neighbors at the previous time step. 
  They are useful models for systems of many 
  identical elements when the dynamics depends only on local
  interactions. 
  Conservative (or ``number-conserving'') cellular automata 
  represent a special class  of CA, 
  in which the {\em sum} of all the states, that are integers, 
  remains constant as the system is iterated.
  This property arises naturally when modeling phenomena such
  as traffic flow (\cite{nagel}), eutectic alloys (\cite{kohyama1,kohyama2}, or the 
  exchange of goods between 
  neighboring individuals. When number-conservation is not apparent
  for the initial
  system, its detection can be interesting by itself, and may help to
  prove dynamical properties.

  Necessary and sufficient conditions for a CA to be number-conserving
  are given in \cite{boccara1} for one dimension and states $\{0,1\}$,
  and in \cite{boccara2} for one dimension and states 
  $\{0,\sdot,q-1\}$;
  a generalization for two and more dimensions
  is found in \cite{durand}. 
  In \cite{moreiratcs} the definition---and the characterizations---are
  extended to allow general sets of states $S\subset \Zset$, and an
  algorithm is given to decide, for any CA, whether its states can
  be relabeled with integer values, so as to make it number-conserving.
  In \cite{morita1} and \cite{morita2} the universality of
  reversible, number-conserving ``partitioned'' CA is proved for
  one and two dimensions, respectively. In \cite{moreiratcs} the
  universality of usual (not partitioned) conservative CA in one dimension
  is proved. In fact, it is shown that any one-dimensional CA can
  be simulated by a conservative CA; this proves the existence of
  {\em intrinsically universal} conservative CA in the sense defined 
  in \cite{oll};
  this notion of universality is stronger than the usual one (the
  ability to simulate universal Turing machines). A construction
  of a logically universal conservative CA in two dimensions
  is given in \cite{imai}; they also construct a 
  self-reproducing model in a two-dimensional conservative CA,
  by embedding in it the well known Langton's loops. Another interesting
  work is found in \cite{durand2}, where the CA classifications
  of K\r{u}rka \cite{kurka1} and Braga \cite{braga1,braga2} are
  intersected and the existence of conservative CA in the resulting
  classes is checked. A recent article by Fuk\'s \cite{fuks2}
  considers probabilistic conservative CA.

  In the articles of Boccara and Fuk\'s (\cite{boccara1,boccara2}) the
  necessary and sufficient condition was used to list and study
  all the conservative CA rules with small neighborhoods and small 
  number of states.
  For all the rules they study, they give a {\em motion 
  representation}: the state of a cell is interpreted
  as the number of particles in it, and the CA rule is interpreted
  as an operator that governs the interaction of these identical, 
  indestructible particles. Fuk\'s \cite{fuks} and Pivato \cite{pivato}
  have independently shown that this interpretation is always
  possible (in the one-dimensional case). In the same
  spirit but with a very general definition of ``particles'', 
  K\r{u}rka \cite{kurka2} has recently considered CA with {\em vanishing} 
  particles.

  For the sake of completeness and to avoid confusions, 
  it is worth mentioning other contexts
  in which particles have been considered. On one hand, there are
  the {\em interacting particle systems} (IPS), with a long
  history in probability theory \cite{ips}, and the {\em lattice gases},
  some of them with associated CA models \cite{boon}; in general,
  they cannot be written as conservative CA. 
  The well-known two-dimensional Margolus CA \cite{margolus} is  
  number-conserving
  and was designed to allow rich interactions of particles; it
  does not fit in the definition given here, because of its
  alternating neighborhood. 
  Particles have been widely
  used in computer graphics \cite{hocknew}, sometimes using
  CA with the neighborhood of Margolus \cite{graph2}.
  The word ``particle'' is also
  used to describe emergent particle-like structures that
  propagate in CA \cite{boccara3,das,hanson,hordijk}; in this last sense,
  it is close to the spirit of our last section, and to that
  in \cite{kurka2}.

  In this article we consider one-dimensional cellular automata;
  Section \ref{sec:defs} gives the necessary definitions and
  reviews (and generalizes) some relevant previous results, while
  Section \ref{sec:defpa} gives our definition of {\em particle
  automata} (PA) as a formalism for motion representation. Section
  \ref{sec:ncca} deals with several issues related to conservative
  CA. First we prove (again) their equivalence with the
  (conservative) PA; then we discuss several
  behaviors that PA may exhibit, showing that some of them
  (like anticipation and global cycles) may be intrinsic to some
  conservative CA. We also consider the special properties
  of state-conservation (where a sensible particle representation
  will recognize each state as a different kind of particle) and
  momentum preservation (which, in spite of being defined
  in terms of the PA, depends only on the conservative CA
  it represents). The main result of the paper is in
  Section \ref{sec:dec}, where we characterize non-increasing
  CA and show how to represent them with particle
  automata. Finally, Section \ref{sec:block} considers
  CA where the states can be relabelled, in a way that 
  depends on the neighbors of a cell, in order to obtain
  a conservative dynamics (and a representation in terms
  of particles).

\section{Definitions and Some Previous Results}
 \label{sec:defs}

 \paragraph*{Cellular automata: }
 A {\em one-dimensional cellular automaton} (CA)
 with set of states $Q=\{0,\sdot,q-1\}$, is any continuous function 
 $F:Q^{\Zset}\to Q^{\Zset}$ which commutes with the shift. It is well known 
 that cellular automata correspond to the
 functions $F$ that can be expressed in terms of a local function: 
 $F(c)_i=f(c_{i+N})$, for all $c\in Q^{\Zset}$, $i\in \Zset$, and
 $N$ a fixed finite subset of $\Zset$, called the {\em neighborhood} of $F$.
 $N$ can always be assumed to be an interval of integers which includes the
 origin, and we write $F(c)_{i+d}=f(c_{i+N})$, with 
 $N=\{0,\sdot,n-1\}$ and $d\in \Zset$ an {\em offset}; rules with the same 
 $f$ but different $d$ will be identical up to a shift.
 It will be useful to define, for $n\in \Nset$ and $Q=\{0,\sdot,q-1\}$,
 \[
  \textrm{\CAa{q}{n}} \; = \; \{f : Q^{\{0,\sdot,n-1\}} \to Q \}
 \]
 Any CA can then be expressed by an element of \CAa{q}{n} for
 some $q$ and $n$, combined with an offset $d$ which tells where the
 image of the neighborhood is placed. The 256 elementary CA, for instance, 
 correspond to \CAa{2}{3}, usually with $d=1$. \CA will denote the
 union of \CAa{q}{n} over all $q$ and $n$.

 A common shorthand notation for cellular automata is the codification used
 by Wolfram~\cite{wolfram}: the code for an element $f\in $\CAa{q}{n} is
 given by 
 \[
  \textrm{Code}(f) \;=\; 
   \sum_{(x_1,\sdot,x_n)\in Q^n} f(x_1,\sdot,x_n) q^{
   \sum_{k=1}^n q^{n-k} x_k}
 \]
 
 \paragraph*{Configurations: }
 An element in $Q^{\Zset}$ is called a {\em configuration}. A configuration
 is said to be {\em finite} if all but a finite number of its components are
 0. A configuration $c$ is said to be {\em periodic} if $c_i = c_{i+p}$, for
 all $i$, for some
 $p\in \Zset$, $p\neq 0$; in this case, $p$ is said to be {\em a period}
 of $c$.

 \paragraph*{Monotone and conserved quantities: }
 Consider a CA $F$ on $\Zset$, and let $C_P$ be the set of all periodic 
 configurations in $\Zset$; for each $c\in C_P$ choose a period $p(c)$.
 Let $\phi$ be a function $\phi:Q^b \to \Rset$, where $b$ is a nonnegative
 integer. Then $\phi$ is said to be a {\em non-increasing additive quantity}
 under $F$ if and only if
  \begin{equation}
   \sum_{k=0}^{p(c)-1} \phi(F(c)_k,\sdot,F(c)_{k+b-1})
    \; \leq \; \sum_{k=0}^{p(c)-1} \phi(c_k,\sdot,c_{k+b-1}),
         \qquad \forall c\in C_P.
     \label{eq:naq}
  \end{equation}
  Similarly, $\phi$ is said to be {\em non-decreasing additive
  quantity} if condition (\ref{eq:naq}) holds with the inequality in the other
  direction. It is easy to see that $\phi$ is non-increasing if and only if
  $-\phi$ is non-decreasing. If $\phi$ is both non-increasing and non-decreasing,
  it is said to be a {\em conserved additive quantity} (in this case,
  (\ref{eq:naq}) holds with an equality sign). We say that $\phi$ is {\em monotone}
  if it is either non-decreasing or non-increasing. In \cite{hattori}
  it is said that ``an additive conserved quantity is a discrete-time analog of
  what we usually call a conserved quantity, such as energy, momentum and charge 
  of a physical system''; the sentence
  can be rephrased for the monotone case.

 \paragraph*{Finitary characterization: }
 The previous definitions consider the addition of a density function
 over a period of a periodic configuration. Another possibility would be
 to consider the sum over finite configurations: we may say that $\phi$
 is a {\em finitely} non-increasing additive quantity if condition (\ref{eq:naq})
 holds for all finite $c$ in $\Zset$, instead of $C_P$, with
 the sums being taken now over the whole $\Zset$. (Here we are assuming that
 $\phi(0,\sdot,0)=0$; if this is not the case, we consider
 $\tilde{\phi}=\phi-\phi(0,\sdot,0)$ instead.)
 It turns out that the two notions are equivalent:

   \begin{thm}[Generalized from \cite{durand}]
   Let $F$ be a CA and $\phi$ be a function $\phi:S^b\to \Rset$. 
	Then $\phi$ is an additive conserved (non-increasing, non-decreasing) 
	quantity for $F$ if and	only if it is an additive 
	finitely conserved (non-increasing, non-decreasing) quantity for $F$.
	\label{teo:fin}
  \end{thm}
  \begin{pf*}{Sketch of the proof.}
    In \cite{durand} the equivalence is proved for conserved
    quantities, in dimension 1, when $\phi:Q\to Q$ is the identity; however,
    their proof includes both the non-increasing and the non-decreasing cases,
    and can be easily extended to the case of a general $\phi$.
    In one direction the proof is trivial: if the condition holds for 
	 all periodic configurations, and $c$ is a finite configuration, then 
	 the condition is shown to hold for $c$ by building a periodic 
	 configuration with blocks that include the non-zero part of $c$. 
	 On the other hand, if the condition is not verified by a
	 periodic configuration with repeated word $w$, then it will be not
	 verified for a finite configuration of the form $\sdot 000 w^N 000 \sdot$,
	 for $N$ large enough: the surplus (or deficit) of the periodic configuration
	 is amplified by the growing $N$, while the only terms that could
	 reduce it (those corresponding to a neighborhood of $0w$ and $w0$)
	 remain fixed. 
	 
	 Notice that the same argument can be also extended
	 to higher dimensions: by repeating an $n$-dimensional pattern enough
	 times, its surplus will be amplified as $N^n$, while the terms
	 corresponding to the border, though not fixed, will grow only as $N^{n-1}$.
  \qed  \end{pf*}

 The following theorem is a useful
 characterization of conserved quantities in one-dimensional CA.

 \begin{thm}[Hattori and Takesue~\cite{hattori}]
  Let $F$ be a one-dimensional CA with local rule $f\in$ \CAa{q}{n}.
  Let $a$ be an arbitrary element in $Q=\{0,\sdot,q-1\}$. 
  Then $\phi:Q^b\to\Rset$ is an additive conserved quantity under $F$ if
  and only if
  \begin{equation}
   \begin{array}{l}
	 \displaystyle
	 \phi_f(x_0,\sdot,x_{b+n-2}) - \phi(x_0,\sdot,x_{b-1})
	 \\
	 \displaystyle
	 \quad =\;
	 \sum_{i=1}^{b+n-2} \{
	  -\phi_f(\underbrace{a,\sdot,a}_i,x_0,\sdot,x_{b+n-2-i})
	  +\phi_f(\underbrace{a,\sdot,a}_i,x_1,\sdot,x_{b+n-1-i})
	  \}
	 \\
	 \displaystyle
	 \quad \;+\;
	 \sum_{i=1}^{b-1} \{
	   \phi( \underbrace{a,\sdot,a}_{b-i},x_0,\sdot,x_{i-1})
	  -\phi( \underbrace{a,\sdot,a}_{b-i},x_1,\sdot,x_i)
	  \}
	 \end{array}
 \end{equation}
  for all $x_0,\sdot,x_{b+n-2} \in Q$,
  where
  \[
   \phi_f(x_0,\sdot,x_{b+n-2}) \;=\; 
	  \phi(f(x_0,\sdot,x_{n-1}),\sdot,f(x_{b-1},\sdot,x_{b+n-2}))
  \]
  \label{teo:hattori}
  \end{thm}

 \paragraph*{Monotone and conservative CA: }
 A cellular automaton is said to be {\em non-increasing} 
 ({\em non-decreasing}, {\em conservative}) if the identity of its 
 set state is a non-increasing ({\em non-decreasing}, {\em conservative})
 quantity for its dynamics. Since the condition depends only
 on the local rule of the CA, and not on the offset,
 we will define \DCAa{q}{n}, \NCCAa{q}{n} and \ICAa{q}{n}
 to be the {\em non-increasing}, {\em conservative}
 and {\em non-decreasing} rules in \CAa{q}{n}, respectively.
 
 Theorem \ref{teo:hattori} implies that $f\in$ \NCCAa{q}{n}
 if and only if, for all $(x_1,\sdot,x_n) \in Q^n$,
   \begin{equation}
    f(x_1,\sdot,x_n) = x_1+\sum_{k=1}^{n-1} 
    \left\{
      f(\underbrace{0,\sdot,0}_{n-k},x_2,\sdot,x_{k+1})
      - f(\underbrace{0,\sdot,0}_{n-k},x_1,\sdot,x_k)
    \right\}
     \label{eq:ncsc}
    \end{equation}

 This characterization, given in~\cite{boccara2}, can be generalized to 
 higher dimensions, though its explicit form becomes hard to write (it
 was done by~\cite{durand}).

 \paragraph*{Some more notation: }

 The letter $q$ will always denote the number of
 states, and the letter $Q$ will denote the set $\{0,\sdot,q-1\}$. 
 With $\bar{0}$ we will denote an infinite sequence of 
 zeroes (thus, $\bar{0}w\bar{0}$ denotes a word $w$ surrounded by
 infinite zeroes).
 For $f\in$ \CAa{q}{n}, we will denote by $f(u/v)$
 the block image of word $u$ when followed by word $v$:
 \[
  f(u/v) \;=\; f(w_0,\sdot,w_{n-1}) \, f(w_1,\sdot,w_n) \, \sdot \, f(w_{|u|-1},\sdot,w_{|u|+n-2})
 \]
 where $w=uv$. Of course, only the $n-1$ first elements in $v$ contribute
 to $f(u/v)$. Furthermore, we denote $f(u)=f(u/\epsilon)$, for any $u$ with
 $|u|\geq n$, where $\epsilon$ is the empty word.

 With this notation, the condition for $f\in$ \CAa{q}{n} to belong to 
 \DCAa{q}{n} can be restated as
 \begin{equation}
  \sum_{k=0}^{|w|-1} f(w/w)_k \; \leq \; \sum_{k=0}^{|w|-1} w_k
  \qquad \forall w\in Q^*
  \label{eq:defdec}
 \end{equation}

 If $g(\bullet)$ is a vector-valued function, we will use the notation 
 $[g(\bullet)]_k$ to refer to its $k$-th component. Finally,
 we will use the function $(\bullet)_+$, defined as
 $(x)_+=x$ for $x>=0$, and $(x)_+=0$ for $x<0$.
 
\section{Particle Automata and Motion Representations}
 \label{sec:defpa}

  A common way to look at conservative CA is through their representation 
  in terms of particles: the state of each cell is interpreted as
  the number of particles contained in it, and 
  a rule is given describing the motion that these
  particles will have, depending 
  on the local context; we will add the possibility of {\em vanishing},
  and will formalize this as {\em particle automata}. 
  
  A particle automaton (PA) with set of states $Q=\{0,\sdot,q-1\}$ will act on
  $Q^\Zset$, like a CA, and like a CA it will be defined by a local rule (or
  rather, by a set of rules) taking as input the states of a local neighborhood,
  which is again some $N$ of the form $\{-\ell,\sdot,r\}$. 
  
  For a local configuration $w=c_{-\ell},\sdot,c_0,\sdot,c_r$ with $c_0 > 0$,
  a function $g_{c_0}$ will give the new positions of the $c_0$ particles at 
  the origin: we have a set $(g_i)_{i=1,\sdot,q-1}$ of functions
  \[
   g_i: Q^\ell \times Q^r \to ( N\cup \dagger )^i
  \]
  where the dagger ($\dagger$) represents the ``vanishing'' option.
  Thus, for a configuration $c\in Q^\Zset$, the $k$-th particle at position
  $i$ (with $1\leq k \leq c_i$) will 
  \[
   \left\{
   \begin{array}{l}
  \textrm{vanish if } [g_{c_i}(c_{i-\ell},\sdot,c_{i-1},c_{i+1},\sdot,c_{i+r})]_k=\dagger \\
  \textrm{move to position } i+[g_{c_i}(c_{i-\ell},\sdot,c_{i-1},c_{i+1},\sdot,c_{i+r})]_k
  \textrm{ otherwise}
   \end{array}
   \right.
  \]
  Since the particles are undistinguishable, the order of
  the components of each $g_i$ is irrelevant; only the number of components
  mapping to each element of $(N\cup \dagger)$ is important.

\begin{figure}[htb]
  \begin{center}
   \resizebox{6cm}{!}{\includegraphics{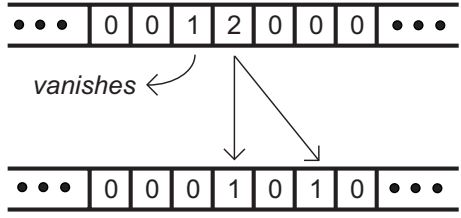}}
	\caption{Example of particle motion}
		\label{fig:itpa}
  \end{center}
\end{figure}

  Some examples may help to make the definition clear. Consider a PA with
  $q$ states that moves all particles one position to the left. It will
  have $\ell=1$, $r=0$ (\ie, $N=\{-1,0\}$), and
  \[
   g_\alpha(\beta) \;=\; (\underbrace{-1,\sdot,-1}_{\alpha}) 
   \qquad
   \textrm{ for all } 0 < \alpha < q, \; 0 \leq \beta < q
  \]
  As a second example, consider a PA with $Q=\{0,1,2\}$, $\ell=1$, $r=2$,
  and $g_1$, $g_2$ such that $g_1(0,2,0)=(\dagger)$ and $g_2(1,0,0)=(0,2)$.
  Figure \ref{fig:itpa} shows its effect on the configuration $\bar{0}12\bar{0}$.

  Thus, a PA $G$ is defined by a tuple $G=(q,N,(g_i)_{i=1,\sdot,q-1})$.
  The global action of $G$ on $c\in Q^\Zset$ is given by
  \[
   G(c)_i \;=\; \min\{\;q-1\;,\; 
	\#\{(j,k): \; j + [g_{c_j}(c_{j-\ell},\sdot,c_{j-1},c_{j+1},\sdot,c_{j+r})]_k = i\}\;\}
  \]
  In other words, $G(c)_i$ is the number of particles arriving at $i$,
  with a maximum of $q-1$; this last condition prevents ``overflows'',
  but, in any case, any rule can always be fixed to avoid needing this
  overflow control, by sending the excess of particles to ``$\dagger$''
  (though this may require an extension of the neighborhood).
   
  We will denote by \PAa{q}{n} the set of all PA with $q$ states
  and neighborhood of size $|N|=\ell+r+1=n$, and by
  \NCPAa{q}{n} the set of conservative PA: those members of \PAa{q}{n}
  such that their functions $g_i$ go to $N^i$ (nothing vanishes),
  and avoid the overflow (\ie, the minimum in the above definition
  is always the right term).
  Notice that unlike the elements of \CAa{q}{n}, an element of \PAa{q}{n}
  is not combined with an offset to define the actual PA.
  
 {\bf Motion Representation: }
 In \cite{boccara1} and \cite{boccara2} Boccara and Fuk\'s give
 {\em motion representations}
 for each of the rules they study; we will follow their notation,
 which is an efficient and intuitive way of expressing a PA. A motion 
 representation is a list
 of specific local configurations of a given a cell, with
 arrows indicating the motion to be performed by the particle(s)
 located in that cell for all these local configurations.
 As a simplification, if for a given local configuration, 
 the particles do not move, then this configuration is not listed.
 Some examples may clarify this point. Consider
 \[
  M_1 \;=\; \{ \overset{\curvearrowright}{10}, 
  \overset{\curvearrowleft}{0011} \}
   \qquad , \qquad
  M_2 \;=\; \{ \overset{\overset{2}{\curvearrowright}}{20}, 
				           \overset{\overset{1}{\curvearrowright}}{21} \}
 \]
 $M_1$, for $q=2$, is read as follows.
 If a particle sees an empty cell on its right, it moves to it. If
 that cell is occupied, but the two cells on its left are empty, it
 moves to the closest one. In any other case, it keeps its current
 position. (The neighborhood in this case is $\{-2,-1,0,1\}$.) 
 Numbers may be added to the arrows when a cell may be
 occupied by more than one particle. This is
 the case in $M_2$, with $q=3$. If two
 particles are in a cell, then one, two, or none of them may move to
 the right, depending on the space available there. Here are two
 more examples:
 \[
  M_3 \;=\; \{ \;
   \overset{k}{ \mbox{ \jumpthreeleft{$\bullet\bullet k$}{3}{3} } }
	  ,  \; k=1,\sdot,q-1  \;\}
    \qquad , \qquad
  M_4 \;=\; \{ \overset{\curvearrowright}{10} \}
  \]
  In $M_3$ the bullets 
  ($\bullet$) are wildcards. In this example, everything moves two
  steps to the left,
  regardless of what the states in the other cells are. 
  If we denote by
  $\sigma$ the {\em shift} (the CA rule such that $f(a_0,a_1)=a_1$, with
  offset 0), then
  $M_3$ represents the CA rule $\sigma^2=\sigma\circ\sigma$.
  Rule $M_4$ (for $q=2$) shows that a particle will move to the right 
  if, and only if, that cell is empty; its effect on $\{0,1\}^\Zset$
  is the same as the elementary CA rule 184, with offset 1.
  The notation is easily extended to include vanishing
  particles, by adding a hat ($\hat{\textrm{ }}$) for each of the particles that
  vanish:
  \[
   M_5 \;=\; \{ \overset{\overset{2}{\curvearrowleft}}{01}, 
		           \overset{\overset{1}{\curvearrowleft}}{11},
		           {2\hat{1}}
					  \}
  \]
  Here the 1's will travel to the left until they meet a 2,
  and then will disappear.

  Note that PA {\em do not} distinguish particles; if $k$
  particles arrive at a same cell, at the next time step the rule says 
  how many of them will move to each neighboring cell, but does
  not say which particles are moving.
  However, particles are, in a sense, distinguishable, since
  we know how many go from each cell to each other cell at each iteration.   
  To say that ``a particle
  moved from $j$ to $i$, while another moved from $i$ to
  $k$'', is not the same as saying 
  ``a particle moved from $j$ to $k$, and another stayed at $i$''. 
  When arbitrarily large groups move, as for instance
  in the shift, we have to assume that some intermediate cells with
  unchanging values {\em are} changing the particles they contain; 
  otherwise, we would need an infinite neighborhood to describe the motion.
  If we implement the system and want to trace the particles throughout 
  the iterations, we need to add some criterion. A sensible choice
  (implicitly applied in \cite{boccara2}) is to keep the order of the 
  particles along the line.

  Both CA and PA are mappings from $S^\Zset$ into itself. We
  say that a CA $F$ is a {\em projection} of a PA $G$ if $F(c)=G(c)$,
  for all $c\in S^\Zset$. The following proposition 
  is straightforward.

  \begin{prop}
   Let $G$ be a PA and $F$ be a projection of $G$. Then $G\in$ \PA
	if and only if $F\in$ \DCA, and $G\in$ \NCPA if and only if 
	$F\in$ \NCCA.
  \end{prop}
 
 \subsection{CA for a given PA}

  \begin{thm}
   For any $G\in$ \PAa{q}{n}, there is a unique CA $F$ with
   local rule $f\in$ \CAa{q}{2n-1}
	which is a projection of $G$. 
	 \label{teo:ca4pa}
  \end{thm}
  \begin{pf*}{Proof.}
   The function $G$ is obviously continuous and shift-commuting. Hence, it
	 may be written as a CA, which is uniquely defined but for the 
	 neighborhood (since the neighborhood size can always be
	 increased). The only thing we have to check
	 is that a neighborhood of size $2n-1$ is enough. 
	 For this note that,
	 in the definition of PA, the state of $c'_i$ is completely determined
    by the particles that will move to $i$. 
	 If we denote by $\{-\ell,\sdot,r\}$ the neighborhood of $G$ ($n=\ell+r+1$),
	 these particles have to be
	 in the cells $\{i-r,\sdot,i+\ell\}$; their behavior, in turn, is 
	 completely 
	 determined by the values in $\cup_{j=i-r}^{i+\ell} 
	 \{j-\ell,\sdot, j+r\}$,
    \ie, by $\{c_{-r-\ell},\sdot,c_{r+\ell}\}$.
  \qed \end{pf*}

  We will denote by $\Pi(G)$ the smallest CA $F$ (with respect to $|N_F|$)
  that is a projection of $G$. If, for $G$, we also take the minimum
  possible neighborhood, then, in most cases, we have $|N_G| < |N_F|$.
  Take, for instance, the motion rule $M_4$ defined above:
  rule 184, a 3-input rule, is the smallest CA matching it. 
  The reason is that an occupied cell must know
  if its particle will leave, \ie, must look to the left, and an
  empty cell must know whether a particle will arrive, \ie, it must
  look to the right. The relation $|N_G| < |N_F|$ is, however, not
  always verified, as shown by the following examples.

  \stepcounter{example}
  {\bf Example \arabic{example}: }
  Consider the PA $G=(2,\{-3,\sdot,2\},g)$ with $g$ described by the
  following motion representation:
  \[
    M_6 \;=\; \{ \mbox{\jumptwoleft{0001}{4}{3}}, 
	 \mbox{\jumponeleft{1001}{4}{3}},
	      \mbox{\jumponeright{110}{3}{1}}, 
			\mbox{\jumponeright{1110}{4}{1}}    
	  \}
  \]
  It may be checked that $\Pi(G)=(\{0,1\},\{-2,\sdot,2\},f)$, with
  $f$ given by code number 3221127170. A way to look at this situation
  is this: from the information given by the occupancy numbers of the
  neighboring cells of an occupied cell, we know that the particle
  will leave it and go to the left by looking at the two cells in that
  direction, but we do not know its precise destination. 
  If we take the viewpoint of the
  destination cells, then we know it, since 
  {\em they do} see the rest of the (PA) neighborhood of the particle.
  In other cases, a cell knows that it will remain occupied, but it 
  does not know if the occupying particle will be the same.
  \qed

  \stepcounter{example}
  {\bf Example \arabic{example}: }
   Here we show a family of PA for which the minimal CA actually requires
	the whole neighborhood allowed by Theorem \ref{teo:ca4pa}. Consider
	the family of PA $G_{\ell,r}$ with $q=3$ described by the
	motion rule
	\[
   M_7 \;=\;
         \{ \,
	 \mbox{ \jumptworight{2u1v00}{5}{2}} \, , \,
	 \mbox{ \jumptwoleft{0u1v2}{5}{2}}  \,
         \}
	 \qquad \textrm{ where } u=2^{\ell-1} \textrm{ and } v=2^{r-1}
	\]
	If we have a local configuration $\alpha u1v0u1v \beta$, which has
	size $2\ell+1+2r$, the next state of the cell in the middle, which
	now contains a 0, depends both on $\alpha$ and $\beta$. Hence, the
	minimal CA has neighborhood $\{-\ell-r,\sdot,\ell+r\}$, while 
	$G$ requires only $\{-\ell,\sdot,r\}$.
	
	If we want
	an example with only two different states, the PA needs to be a bit
	more complicated; one possibility is described by the motion rule
	\[
   M_8 \;=\;
        \{ \,
	 \mbox{ \jumptworight{00u1v0}{6}{3}} \, , \,
	 \mbox{ \jumpthreeleft{11u1v0}{6}{3}} \, , \,
	 \mbox{ \jumptworight{110}{3}{0} } \,
          \}
	 \quad \textrm{ where } u=0^{\ell-2} \textrm{ and } v=0^{r-1}
	\]
	\qed

 \section{Particles for Conservative CA}
  \label{sec:ncca}

 \subsection{PA for a given conservative CA}

  \begin{thm}
   For each conservative CA $F$ with local rule $f\in$ \NCCAa{q}{n},
	there exists $G\in$ \NCPAa{q}{2n-1}, such that $\Pi(G)=F$.
  \label{thm:pa4ca}
  \end{thm}
  \begin{pf*}{Proof.}
   This result was independently proved both by Fuk\'s~\cite{fuks}
   and Pivato~\cite{pivato}. Though very different, all these
   first proofs (including an unpublished one by ourselves) are 
   rather long; the idea, however is very simple, and we sketch it
   here. In addition, Theorem \ref{thm:eqdca} in Section \ref{subs:pa4dca} will
   generalize it, providing yet another proof for the theorem.

   Let $\{-\ell,\sdot,r\}$ be the neighborhood of $F$. We define $G$
   with $N_G=\{-(\ell+r),\sdot,(\ell+r)\}$; for all 
   $w=w_{-\ell-r},\sdot,w_{\ell+r}$$\in Q^{2n-1}$ with $w_0>0$, we have to define
   the new positions $g_{w_0}(w_{-\ell-r},\sdot,w_{-1},w_1,\sdot,w_{\ell+r})$. 
   This is done by computing the image
   of the configuration $c=\bar{0}w\bar{0}$, with $w_0$ at the origin, 
   and matching the particles in the image $c'$ with the particles in $c$, from left to right.

   We claim that the new positions of the $w_0$ particles at the origin are in
   $\{-r,\sdot,\ell\}$. Suppose that one of them goes to a position
   to the left of $-r$: this means that the particles in $(c_i)_{i<0}$
   did not match all the particles in $(c'_i)_{i<-r}$. 
   Since these $c'_i$ depend only on the values of $(c_i)_{i<0}$,
   this would contradict $f\in$ \NCCA for the configuration 
   $\bar{0}w_{-\ell-r}\sdot w_{-1}\bar{0}$. The symmetric argument
   shows that no particle from the origin moves to the right of position $\ell$.

   Now take any $u\in Q^*$, and define
   $c=\bar{0}uw\bar{0}$ and $c'=F(c)$. As before, 
   we match the particles in $c$ and $c'$, from left to right.
   Notice that we have added $\sum_{i=0}^{|u|-1} u_i$ particles in the
   preimage; since $f\in$ \NCCA, we have also added the same number of
   particles in the image, and they must be to the left of $-r$ (the rest 
   of the image has not changed). Hence, the movement
   of the $w_0$ particles at the origin is the same as before, and we see
   that the effect of applying $G$ on any finite configuration $c$ is the
   same as the effect of matching the particles with those from the
   image of $c$ through $F$. We conclude that $F$ and $G$ have the
   same effect on $c$, and $F=\Pi(G)$.
  \qed \end{pf*}

  \begin{cor}
   \label{cor:regular}
   In Theorem \ref{thm:pa4ca}, if the CA $F$ has offset $\ell$ (and hence 
   neighborhood $\{-\ell,\sdot,r\}$,
   with $r=n-1-\ell$), then $G$ has neighborhood $\{-(\ell+r),\sdot,(\ell+r)\}$,
   and for each $i=1,\sdot,q-1$, $g_i(Q^{2\ell+2r})\subset \{-r,\sdot,\ell\}^i$
   (\ie, the particles move only to $\{-r,\sdot,\ell\}$).
  \end{cor}

  \refstepcounter{example}
  \label{example:detabla}
  {\bf Example \arabic{example}: }
  If we take $f\in$ \NCCAa{2}{5} with code \#288146448 (see Table \ref{tab:regla})
  and offset 2, we obtain
  a PA with a motion representation given by
  \[
   M_9 \;=\;
   \{
	 	  \mbox{\jumptworight{1$\bullet$00}{4}{0}},
	    \mbox{\jumptworight{1001}{4}{0}},
	    \mbox{\jumptworight{1010}{4}{0}},
	    \mbox{\jumponeright{1011}{4}{0}},
	    \mbox{\jumponeright{1101}{4}{0}}
	\}
  \]
  If we set the offset to 0, the result is
  \[
   M_{10} \;=\;
   \{
	   \mbox{\jumponeleft{$\bullet$1011}{5}{1}},
	   \mbox{\jumponeleft{$\bullet$1101}{5}{1}},
	   \mbox{\jumptwoleft{$\bullet\bullet$111}{5}{2}}
	\}
  \]

  \refstepcounter{example}
  \label{example:321}
  {\bf Example \arabic{example}: }
  If we take $f\in$ \NCCAa{2}{4} with code \#49024 (see Table \ref{tab:regla2})
  and use offset 1, the resulting PA has motion representation
  \[
   M_{11} \;=\;
   \{
		\mbox{\jumponeright{1$\bullet$0}{3}{0}},
		\mbox{\jumponeright{100}{3}{0}}
	\}
  \]\qed
  
 \begin{table}[htb]
  \begin{center}
  \begin{scriptsize}
  \begin{tabular}{||c|c||c|c||c|c||c|c||} \hline \hline
    00000 & 0 & 01000 & 0 & 10000 & 1 & 11000 & 1 \\ \hline
    00001 & 0 & 01001 & 0 & 10001 & 1 & 11001 & 1 \\ \hline
    00010 & 0 & 01010 & 0 & 10010 & 1 & 11010 & 0 \\ \hline
    00011 & 0 & 01011 & 1 & 10011 & 1 & 11011 & 1 \\ \hline
    00100 & 0 & 00100 & 0 & 10100 & 1 & 11100 & 0 \\ \hline
    00101 & 0 & 00101 & 1 & 10101 & 1 & 11101 & 1 \\ \hline
    00110 & 0 & 00110 & 0 & 10110 & 0 & 11110 & 0 \\ \hline
    00111 & 1 & 00111 & 1 & 10111 & 1 & 11111 & 1 \\ \hline
    \hline
  \end{tabular}
  \end{scriptsize}
  \caption{Lookup table for rule \#2881464448 in \NCCAa{2}{5}}
   \label{tab:regla}
  \end{center}
 \end{table}

 \begin{table}[htb]
  \begin{center}
  \begin{scriptsize}
  \begin{tabular}{||c|c||c|c||c|c||c|c||} \hline \hline
    0000 & 0 & 0100 & 0 & 1000 & 1 & 1100 & 1 \\ \hline
    0001 & 0 & 0101 & 0 & 1001 & 1 & 1101 & 1 \\ \hline
    0010 & 0 & 0110 & 0 & 1010 & 1 & 1110 & 0 \\ \hline
    0011 & 0 & 0111 & 1 & 1011 & 1 & 1111 & 1 \\ \hline
    \hline
  \end{tabular}
  \end{scriptsize}
  \caption{Lookup table for rule \#49024 in \NCCAa{2}{4}}
   \label{tab:regla2}
  \end{center}
 \end{table}

 \subsection{On Some Behaviors of PA}

  The theorem in the preceding section shows how to construct, 
  for a given conservative CA $F$, a conservative PA $G$ 
  such that $F=\Pi(G)$. We
  will call this the {\em canonical} PA for $F$, since it is
  the only one that preserves the order of the particles
  along the line. However, it is not the only PA that
  matches $F$ (in fact, there are infinite PA matching any
  given CA). For this reason, we shall discuss some behaviors 
  that a PA may exhibit, and whether or not they can be 
  intrinsic to certain CA; this may be relevant in the
  applications.

  {\bf Local cycles: }
  For a PA and a given configuration, we say that there is
  a {\em local $m$-cycle} in the iteration if there is a chain 
  of particles 
  $p_0, p_1,\sdot,p_m$, with $p_0=p_m$, all located in
  different cells, 
  such that, for $i=0,\sdot,m-1$, each $p_i$ moves to the cell
  occupied previously by particle $p_{i+1}$.
  This behavior may be unwanted if the rule is supposed to express
  the motion of indistinguishable elements. 
  The following motion representation has a 3-cycle; less trivial
  local cycles can be found in other PA (and it is even possible to construct
  PA with local cycles of arbitrary length).
  \[
   M_{12} \;=\;
    \{
	     \mbox{\jumponeright{01110}{5}{1}}, \mbox{\jumponeright{01110}{5}{2}},
		   \mbox{\jumptwoleft{01110}{5}{3}}
	  \}
  \]

  {\bf Order preservation: }
  We say that a PA preserves the order if there is no configuration 
  in which a particle moves from position $i_0$ to position $i_1$,
  while another moves from $j_0$ to $j_1$,
  with $i_0 < j_0 \leq j_1 < i_1$,
  or $i_1<j_1\leq j_0<i_0$. This behavior may
  be wanted, for instance, when modeling cars moving
  on a one-lane road. Order-preserving 
  PA do not admit local cycles.

  {\bf Anticipation: }
  We say that a PA with states $\{0,\sdot,q-1\}$ exhibits {\em anticipation}, if for some
  configuration a cell in a state $s$ receives, from
  other cells, a number of particles $t$ such that $t+s>q-1$ 
  (as if, as a result of the rule,
  the neighboring cells ``knew'' that some particles will leave
  the cell). For $q=2$, anticipation is what happens when a particle
  moves to a currently occupied cell.
  In the case of highway traffic, for instance, most drivers usually 
  move anticipating the motion of the cars ahead (assuming that they 
  are not going to stop), and this is considered in the models. 
  An important case in which anticipation may be unwanted is the 
  situation in which movement is optional: the particles may decide to
  stay in their place or to move, but if they move, they must
  obey the PA rule. In this situation, an anticipating rule would cause
  collisions (or overflows, if $q>2$).

  {\bf Global cycles: }
  We say that there is a {\em global cycle} if there is a chain of 
  particles 
  $\{p_i\}_{i\in\Nset}$, located at different positions, such 
  that, for all $i$,
  particle $p_i$ moves to the cell previously occupied by
  particle $p_{i+1}$.
  The reason to call this a
  ``cycle'' is the following. By removing local cycles, the chain may be
  assumed to approach $\infty$ (or $-\infty$) as $m\rightarrow \infty$. 
  If the 
  configuration is---spatially---periodic of period $p$, then we 
  may identify 
  it with the torus $\Zset_p$, and the chain is in fact a cycle. If the 
  configuration is not periodic, as we follow the chain we will at 
  some point
  find a repetition (due to the finite number of possibilities) of a 
  block of length larger than $|N|$. At this point we may cut the 
  part of the configuration starting with 
  the block and ending with it, and repeat it to produce a periodic 
  configuration
  with a cyclic replacement. A trivial example of global cycles is the
  PA that just shifts the configuration, as motion rule $M_3$ given above.
  
  A special case are global cycles of anticipatory motion (which is
  always the case for global cycles when $q=2$). Such
  cycles may be unwanted if we are modeling agents with local information,
  and we do not want them to use the information ``I am on a torus'',
  or ``I am in an infinite queue''.

 \begin{thm}
  \label{thm:intrinsic}
  For any conservative CA $F$ there exists a $G\in $ \NCPA such that 
  $\Pi(G)=F$ and which preserves the order (and hence, has no local cycles). 
  On the other hand, there do exist rules in \NCCA for which anticipation 
  and global cycles are intrinsic: with any offset, 
  they are not the projection of any PA without these features.
 \end{thm}
  \begin{pf*}{Proof.}
  The first part follows from the construction in Theorem \ref{thm:pa4ca}:
  the canonical PA preserves the order.
  For the second, we just need to exhibit a CA for which the claimed
  property is true. 

  Consider a CA $F$ with the local rule $f\in$ \NCCAa{2}{5} 
  of Example \ref{example:detabla}.
  We will show now that any $G\in$ \NCPA such
  that $\Pi(G)=F$ must have global cycles (and, in particular, anticipation),
  for any offset.
  Let $G\in$ \NCPAa{q}{n} be such that $\Pi(G)=F$. Consider 
  a configuration
  \[
   \sdot 00000.111111\sdot 1111111,0000000\sdot
  \]
  where the sequence of 1's is longer than $2n$; the dot and the
  comma are there for reference. The image of this configuration,
  assuming an offset 0, is
  \[
   \sdot 000011.111111\sdot 1110011,0000000\sdot
  \]
  If the particles in the middle of the configuration {\em are moving}, 
  then they are moving without seeing any 0's; hence, the particles in the 
  configuration $\bar{1}$ (where all states are 1) would also be moving, 
  and that would be a global cycle. On the other hand, if the particles 
  in the middle are {\em not moving}, then there are two particles that 
  are moving somehow from one end of the region of 1's to the other,
  which is a contradiction, since the region is larger than the
  neighborhood of the PA.

  For any other choice of the offset, we obtain the same situation, 
  except for an offset of 2.
  But in that case, we can consider the configuration
  \[
   000000.1010101010 \sdot 010101,00000
  \]
  whose image is
  \[
   000000.0010101010 \sdot 010101,01000
  \]
  and produces the same result as above: for any PA, there is an 
  anticipation to the right, and it allows a global cycle.
  \qed \end{pf*}

  If anticipation and global anticipatory cycles may be intrinsic 
  to a rule in \NCCA, then it is natural to ask about the
  decidability of these properties: given a rule $f\in$ \NCCA,
  can we decide if, for some offset, there does exist a PA
  without anticipation (and/or without global anticipatory cycles) 
  from which the CA is the projection? The answer is not trivial.
  It is easy to check these properties on a given PA; however,
  to check them on a given CA rule, we must consider {\em all} the
  possible PA for which the CA is the projection. Consider
  the CA of Example \ref{example:321}: the canonical PA given
  by Theorem \ref{thm:pa4ca} shows anticipation, but a non-anticipating
  PA with the same projection does exist, and is described by the motion
  representation
  \[
   M_{13} \;=\;
	\{
	 \mbox{\jumptworight{110}{3}{0}}
	 \mbox{\jumponeright{010}{3}{1}}
	\}
  \]
  Hence, to find a non-anticipating PA we may need to drop the
  condition of order preservation. In fact, the situation is even 
  worse. The canonical PA of Theorem \ref{thm:pa4ca} has a nice 
  feature, stated in Corollary \ref{cor:regular}: a particle moves 
  always to a cell that ``sees'' it in its CA neighborhood.
  As we shall see in the next sections, we can ask for the same feature
  when obtaining PA for monotone CA and for state-conserving
  CA; it seems to be ``natural'', and some authors have taken for
  granted that only motion rules with this property need to be considered.
  But to avoid anticipation, we may need to drop this condition too.
  It is violated in the PA described by 
  $M_{13}$, and it can be shown that anticipation
  cannot be avoided without violating it for the local rule of Example
  \ref{example:321}, combined with {\em any} offset. Thus, we must consider
  a rather large set of PA for a given CA rule in order to decide
  if anticipation can be avoided or not.

  \begin{thm}
  For a given $f\in$ \NCCA, it may be decided whether an offset
  $\ell$ exists for which the CA defined by $f$ and $\ell$
  is the projection of some PA without anticipation. If the answer
  is positive, that PA can be found.
  \label{thm:final}
  \end{thm}
  \begin{pf*}{Proof.}
   First we must notice that for a given $f\in$ \NCCA, there is at most
   one offset $\ell$ for which a non-anticipating PA can exist. This
   was already seen in the proof of Theorem \ref{thm:intrinsic} for the rule 
   of Example \ref{example:detabla}: when we evaluated a configuration
   of the form $\bar{0}1^m\bar{0}$, with an arbitrarily large $m$, an
   offset had to be imposed to prevent the movement of particles from
   one extremity of the 1's to the other. For the general case,
   let us suppose that there is a non-anticipating PA $G$ such that
   its projection is $f$ with offset $\ell$ (and hence neighborhood
   $\{-\ell,\sdot,r\}$, with $r=n-1-\ell$). Then we may assume that $G$
   has a neighborhood $\{-L,\sdot,R\}$ with $L \geq \ell$ and $R \geq r$.
   Consider a configuration $c=\bar{0}(q-1)^M\bar{0}$, with $M$ arbitrarily
   large. We have the situation depicted in Table \ref{tab:mono}.

 \begin{table}[htb]
  \begin{center}
  \begin{scriptsize}
  \begin{tabular}{||c||c|c|c|c|c|c|c||} \hline \hline
     &  A        & B                             & C                & D                 & E             & F                             & G         \\ \hline
$c$  & $\bar{0}$ & $0^r (q-1)^\ell$              & $(q-1)^{L-\ell}$ & $(q-1)^{M-(L+R)}$ & $(q-1)^{R-r}$ & $(q-1)^r0^\ell$               & $\bar{0}$ \\ \hline
$c'$ & $\bar{0}$ & $f(0^{\ell+r}(q-1)^{\ell+r})$ & $(q-1)^{L-\ell}$ & $(q-1)^{M-(L+R)}$ & $(q-1)^{R-r}$ & $f((q-1)^{\ell+r}0^{\ell+r})$ & $\bar{0}$ \\ \hline
    \hline
  \end{tabular}
  \end{scriptsize}
  \caption{Iteration on $\bar{0}(q-1)^M\bar{0}$}
   \label{tab:mono}
  \end{center}
 \end{table}

   The PA must keep the particles in region D fixed: they only see cells in 
   state $q-1$ around them. If they move, then we would have anticipation (and a global
   anticipatory cycle) for the configuration where all cells are in state $q-1$. Furthermore,
   the particles in regions B and C cannot move to region D (this would imply anticipation),
   nor to regions E, F and G (since $M$ is arbitrarily large). Hence, $G$ must match
   the $L(q-1)$ particles from regions B and C with the particles that $c'$ has in
   regions B and C, and this requires
   \[
    L(q-1) \;=\; (L-\ell)(q-1) + \sum_{i=1}^{n-1} 
    f(\underbrace{0\sdot 0}_{i}\underbrace{(q-1)\sdot (q-1)}_{n-i})
   \]
   which implies
   \begin{equation}
    \label{eq:elequ}
    \ell \;=\; \frac{1}{q-1} \sum_{i=1}^{n-1} 
    f(\underbrace{0\sdot 0}_{i}\underbrace{(q-1)\sdot (q-1)}_{n-i})
   \end{equation}
   Thus an offset is imposed for the existence of a non-anticipating PA. If the
   right side of (\ref{eq:elequ}) is not an integer, then we are done with $f$:
   any PA will have to show anticipation, and, moreover, a global anticipatory cycle
   (since the anticipation takes place through the arbitrarily large region D, and
   hence, will also take place in the configuration with all cells in state $q-1$).

  We will attempt to define a non-anticipating PA $G$ with neighborhood 
  $\{-L,R\}$, where $L=L_1+\ell$, $R=R_1+r$, and $L_1, R_1$ are ``large'' numbers that 
  will be precised below. If the attempt fails, we will show that a global 
  anticipatory cycles must exist.

  The construction of $G$ is inspired by the proof of Theorem \ref{thm:pa4ca}. For
  every sequence $w=w_{-L},\sdot,w_R\in Q^{L+R+1}$, we will consider the configuration
  $c=\bar{0}w\bar{0}$ with $w_0$ at the origin and will use it to define the new
  positions of the $w_0$ particles. As before, we write $c'=F(c)$, where $F$ is the
  CA defined by $f$ and the (now fixed) offset $\ell$.

  If $G$ is to be non-anticipating, then there are certain cells were we already now
  that some particles cannot move. Without anticipation, the cell at position $i$
  can receive at most $q-1-c_i$ particles from other cells. If $c'_i > q-1-c_i$, then
  at least $c'_i-(q-1-c_i)$ particles in position $i$ must
  stay there, without moving. For every $i\in \Zset$, we define
  \[
    b_i \,=\, (c'_i+c_i-q+1)_+
    \qquad
    a_i \,=\, c_i-b_i
    \qquad
    d_i \,=\, c'_i-b_i
  \]
  Thus, $(b_i)_{i\in \Zset}$ are the numbers of particles fixed at the 
  different positions, $(a_i)_{i\in \Zset}$ contains 
  the numbers of particles in the preimage which need to be associated to some
  particles in the image, and $(d_i)_{i\in \Zset}$ has the numbers of particles
  in the image which need to be associated with some in the preimage.
  Clearly, $a_i=b_i=c_i=0$ for $i\notin \{-L_1-\ell,R_1+r\}$, and $d_i=0$ for
  $i\notin \{-L_1-\ell-r,R_1+r+\ell\}$.
  As in Theorem \ref{thm:pa4ca}, we associate the two sets of particles, $(a_i)_{i\in \Zset}$
  and $(d_i)_{i\in \Zset}$, from
  left to right. The PA is thus defined: the particles at the origin have found their
  new locations, possibly fixing some of them ($b_0$), and associating the rest to
  the available particles in the image. Notice that in general this PA does not
  preserve the order of the particles.

  Suppose now that (for at least some $w$), a particle from the
  origin is sent to a position outside of $\{-L_1,R_1\}$. We consider the case in which
  it is associated to a position to the left of $-L_1$ (the other case is symmetric); this means
  that the non-fixed particles to the left of 0 in $c$ were not enough to match all 
  the non-fixed particles to the left of $-L_1$ in $c'$, \ie
  \begin{eqnarray}
  & &
   \sum_{i=-L_1-\ell}^{-1} a_i < \sum_{-L_1-\ell-r}^{-L_1-1} d_i  
   \nonumber   \\
  & \iff &
   \sum_{i=-L_1}^{-1} a_i < \sum_{-L_1-\ell-r}^{-L_1-1} d_i  - \sum_{i=-L_1-\ell}^{-L_1-1} a_i
   \nonumber   \\
  & \iff &
   \sum_{i=-L_1}^{-1} a_i < \sum_{-L_1-\ell-r}^{-L_1-1} d_i  + \sum_{i=-L_1-\ell}^{-L_1-1} b_i
                          - \sum_{i=-L_1-\ell}^{-L_1-1} b_i - \sum_{i=-L_1-\ell}^{-L_1-1} a_i
   \nonumber   \\
  & \iff &
   \sum_{i=-L_1}^{-1} a_i 
    < \sum_{-L_1-\ell-r}^{-L_1-1} c'_i - \sum_{i=-L_1-\ell}^{-L_1-1} c_i 
    \label{eq:cond1}
  \end{eqnarray}

  On the other hand, 
  \begin{equation}
   \sum_{-L_1-\ell-r}^{-L_1-1} c'_i 
   \;\leq\;
   \sum_{-L_1-\ell-r}^{-L_1-1+r} c_i
   \;\leq\;
   r(q-1) + \sum_{-L_1-\ell-r}^{-L_1-1} c_i
    \label{eq:cond2}
  \end{equation}
  (otherwise, the configuration $\bar{0}c_{-L_1-\ell},\sdot,c_{-L_1-1+r}\bar{0}$
  would contradict $f\in$ \NCCA). Combining (\ref{eq:cond1}) and (\ref{eq:cond2})
  we obtain
  $
  \sum_{i=-L_1}^{-1} a_i < r(q-1)
  $; thus, by choosing $L_1\geq r(q-1)L_2$ (where $L_2$ has still to be chosen) 
  we can guarantee that there will be an interval $I=\{s,\sdot,s+L_2-1\}
  \subset \{-L_1,\sdot,-1\}$, such that $a_i=0$ for all $i\in I$, \ie,
  all particles in this region are fixed.
  
  Now, we define $t=2\max\{\ell,r\}$, and choose $L_2=(t+1)q^t$. In this way,
  we guarantee the existence of a word $u$ of length $t$ that occurs at least twice,
  without overlap, in $c_s,\sdot,c_{s+L_2-1}$. We can write it as $u=u_1 u_2$,
  with $|u_1|=|u_2|=\max\{\ell,r\}$. Let $-T$ and $-T'$ be the positions to the left of the
  first occurrence of $u$, and to the right of its second occurrence, respectively. 
  If we write $x=c_{-L_1-\ell},\sdot,c_{-T}$, $x'=c'_{-L_1-\ell-r},\sdot,c'_{-T}$,
  $y=c_{-T'},\sdot,c_{R_1+r}$, and $y'=c'_{-T'} ,\sdot, c'_{R_1+r+l}$, we can rewrite
  $c$ as $c=\bar{0} x u_1 u_2 u_3 u_1 u_2 y \bar{0}$ for some $u_3$, and 
  we have that
  \[
   F(c) = F(\bar{0} x u_1 u_2 u_3 u_1 u_2 y \bar{0}) = \bar{0} x' \tilde{u}'_1 u'_2 u'_3 u'_1 \tilde{u}'_2 y' \bar{0}
  \]
  If we define $c^M=\bar{0} x u_1 (u_2 u_3 u_1)^M u_2 y \bar{0}$, for an arbitrary 
  positive integer $M$, we obtain that
  \[
   F(c^M) = F(\bar{0} x u_1 (u_2 u_3 u_1)^M u_2 y \bar{0}) 
   = \bar{0} x' \tilde{u}'_1 (u'_2 u'_3 u'_1)^M \tilde{u}'_2 y' \bar{0}
  \]
  The situation is similar to the first paragraphs of this proof. 
  Since $M$ is arbitrarily large, we have an arbitrarily large region were 
  particles cannot be moved by {\em any} non-anticipating PA. If we choose
  $M$ such that $|u u_3|M > L'+R'+1$, where $\{-L',\sdot,R'\}$ is the neighborhood
  of any candidate non-anticipating PA, we see that the PA will have to match
  the particles of $x u_1$ with those of $x'\tilde{u}'_1$. This requires an equal 
  number of particles in both of them, \ie,
  \begin{eqnarray*}
  & & \sum_{i=-L_1-\ell}^{-T+t} c_i = \sum_{i=-L_1-\ell-r}^{-T+t} c'_i
  \\
   & \iff &
   \sum_{i=-L_1-\ell}^{-L_1-1} c_i + \sum_{i=-L_1}^{-T+t} a_i + \sum_{i=-L_1}^{-T+t} b_i= 
   \sum_{i=-L_1-\ell-r}^{-L_1-1} c'_i + \sum_{i=-L_1}^{-T+t} c'_i 
  \\
   & \iff &
   \sum_{i=-L_1}^{-T+t} b_i - \sum_{i=-L_1}^{-T+t} c'_i = 
   \sum_{i=-L_1-\ell-r}^{-L_1-1} c'_i - \sum_{i=-L_1-\ell}^{-L_1-1} c_i - \sum_{i=-L_1}^{-T+t} a_i
  \\
   & \Rightarrow &
   \sum_{i=-L_1}^{-T+t} b_i - \sum_{i=-L_1}^{-T+t} c'_i \geq
   \sum_{i=-L_1-\ell-r}^{-L_1-1} c'_i - \sum_{i=-L_1-\ell}^{-L_1-1} c_i - \sum_{i=-L_1}^{-1} a_i
   > 0
  \end{eqnarray*}
   where the last inequality uses (\ref{eq:cond1}). We have arrived to
   a contradiction, since $b_i \leq c'_i$, for all $i$.

   From all the previous discussion, we see that our construction of $G$,
   with neighborhood $\{-L_1-\ell,\sdot,R_1+r\}$, $L_1=r(q-1)(t+1)q^t$,
   $R_1=\ell(q-1)(t+1)q^t$, will move the particles from a position $i$
   to $\{i-L_1,\sdot,i+R_1\}$; if not, then $f$ does not admit a non-anticipating
   PA, and it exhibits global anticipatory cycles, with any offset.

   We still have to show that $G$ is really non-anticipating, and that $F$ is
   its projection. For the first fact, we have to notice that the cell
   at position $i$ is receiving at most $d_i$ particles (it may receive less
   than $d_i$, since the assignment of non-fixed particles may fix some of
   them). For $G$ to be non-anticipating we need 
   \[
    d_i \leq q-1-c_i
    \iff
    c'_i-b_i \leq q-1-c_i
    \iff
    c'_i +c_i - q + 1 \leq b_i
   \]
   which follows from the definition of $b_i$.

   To show that $F$ is a projection of $G$, we follow again the scheme of Theorem \ref{thm:pa4ca},
   and consider the effect of adding an arbitrary word $u$ to the left of
   $w$. Instead of $c=\bar{0}w\bar{0}$, we take now $\tilde{c}=\bar{0}uw\bar{0}$,
   and proceed as before: we define $\tilde{b_i}$, $\tilde{a_i}$, $\tilde{d_i}$,
   fix $\tilde{b_i}$ particles at position $i$, and associate the ``free'' particles
   from left to right. We will show that the destination of the $w_0$ particles
   at the origin is exactly the same as before; this implies
   that applying $G$ to a finite configuration $c$
   has the same effect of matching the particles with those from $F(c)$, and we conclude that $\Pi(G)=F$.

   In order to show that the destination of the particles at the origin has
   not changed, we will show that the number of ``free particles'' sent from
   $\{-L_1,\sdot,-1\}$ to positions to the left of $-L_1$
   is not changed. Since for $i >= -L_1$ we have not only $\tilde{c_i}=c_i$, 
   but also $\tilde{c'_i}=c'_i$ (and hence $\tilde{b_i}=b_i$, $\tilde{a_i}=a_i$,
   $\tilde{d_i}=d_i$), this implies that the destinations of the particles
   at the origin remain the same.

   The number of ``free particles'' sent from positions 
   $\{-L_1,\sdot,-1\}$ to positions to the left of $-L_1$ is equal to the difference between the
   number of particles to the left of $-L_1$ in the image, and the number of particles
   to the left of $-L_1$ in the preimage. Thus, we have to show that
   \begin{eqnarray}
    & & 
    \sum_{i=-L_1-\ell-r}^{-L_1-1} c'_i - \sum_{i=-L_1-\ell}^{-L_1-1} c_i 
     \;=\; 
    \sum_{i=-L_1-\ell-|u|-r}^{-L_1-1} \tilde{c}'_i - \sum_{i=-L_1-\ell-|u|}^{-L_1-1} \tilde{c}_i
    \nonumber \\
    & \iff & 
    \sum_{i=-L_1-\ell-r}^{-L_1-1} c'_i  
     \;=\; 
    \sum_{i=-L_1-\ell-|u|-r}^{-L_1-1} \tilde{c}'_i - \sum_{i=0}^{|u|-1} u_i
    \label{eq:ultima1}
   \end{eqnarray}
   Since $F$ is conservative, and since $\tilde{c}'_i=c'_i$ for $i>=-L_1$, we
   have
   \begin{equation}
    \sum_{i=-L_1-\ell}^{R_1+r} w_i 
    = \sum_{i=-L_1-\ell-r}^{-L_1-1} c'_i + \sum_{i=-L_1}^{R_1+r+l} c'_i
    \label{eq:ultima2}
   \end{equation}
   and
   \begin{equation}
    \sum_{i=0}^{|u|-1} u_i + \sum_{i=-L_1-\ell}^{R_1+r} w_i 
    = \sum_{i=-L_1-\ell-|u|-r}^{-L_1-1} \tilde{c}'_i + \sum_{i=-L_1}^{R_1+r+l} \tilde{c}'_i
    = \sum_{i=-L_1-\ell-|u|-r}^{-L_1-1} \tilde{c}'_i + \sum_{i=-L_1}^{R_1+r+l} c'_i
    \label{eq:ultima3}
   \end{equation}
   We obtain (\ref{eq:ultima1}) from (\ref{eq:ultima2}) and (\ref{eq:ultima3}).
  \qed \end{pf*}

 \begin{cor}
  The intrinsic presence of anticipation and the intrinsic presence of
  global anticipatory cycles are equivalent properties for conservative CA rules.
 \end{cor}
 \begin{pf*}{Proof.}
  One direction is trivial, since global anticipatory cycles include anticipation. The other
  follows from Theorem \ref{thm:final}: if the procedure fails to give a non-anticipating PA,
  then a global anticipatory cycle occurs.
  \qed \end{pf*}

  A comment about the feasibility of finding a non-anticipating PA is due. Even if Theorem
  \ref{thm:final} asks, in principle, for the evaluation of a huge number of con\-fi\-gu\-ra\-tions
  ($(q-2)(q-1)^{(n-1)(q-1)(t+1)q^t+n-1}$, with $t=2\max\{\ell,r\}$), more practical
  implementations are possible: if we apply the procedure of the theorem to a configuration
  $\bar{0}w_{-\alpha},\sdot,w_0,\sdot,w_{\beta}\bar{0}$, with $\alpha \geq \ell$ and $\beta \geq r$,
  and we find that the $w_0$ particles at the origin move to positions in $\{-\alpha+\ell,\sdot,\beta-r\}$,
  then the argument contained in the last part of the proof holds, and we do not need to
  evaluate the configurations of the form $u w' v$: the movement of these $w_0$ will be the same in all cases.

  \subsection{State-conserving CA}

  Some CA satisfy a condition stronger that number-conservation: 
  they conserve
  the {\em number of cells in each state}, throughout the iterations. 
  We say
  that $f\in$ \CAa{q}{n} is {\em state-conserving} if it 
  verifies
 \begin{equation}
  \sum_{k=0}^{|w|-1} \delta_\alpha(f(w/w)_k) \; = \; \sum_{k=0}^{|w|-1} \delta_\alpha(w_k)
  \qquad \forall w\in Q^*, \alpha\in Q
  \label{eq:defscca}
 \end{equation}
  where $\delta_\alpha(x)$ is 1 for $x=\alpha$ and 0 otherwise. 
  It is plain to see that state-conserving CA are conservative,
  and that all elements in \NCCAa{2}{n} are state-conserving.
  State-conserving CA may be characterized as follows:

  \begin{prop}
  A necessary and sufficient condition for $f\in$ \CAa{q}{n}
  to be state-conserving is that, for all $\alpha, x_1,\sdot,x_n \in Q$,
   \[
	 \begin{array}{l}
	 \displaystyle
    \delta_\alpha(f(x_1,\sdot,x_n))\\
	 \displaystyle
	  \quad =\;
	  \delta_\alpha(x_1)+\sum_{k=1}^{n-1} \delta_\alpha(f(\underbrace{0,
	  \sdot,0}_{n-k},x_2,\sdot,x_{k+1}))
      - \delta_\alpha(f(\underbrace{0,\sdot,0}_{n-k},x_1,\sdot,x_k))
	 \end{array}
    \]
  \label{teo:scca}
  \end{prop}
  \begin{pf*}{Proof.}
  For each $\alpha \in Q$, we apply Theorem \ref{teo:hattori} to
  the density function $\delta_\alpha$.
  \qed \end{pf*}
  
  Notice that $\delta_\alpha$ is a non-linear function. Thus, unlike
  the characterization of \NCCA in \ref{eq:ncsc}, this
  characterization cannot be used to create a linear system
  whose solutions would give all the CA satisfying the condition; however,
  it may be used as a test to look for state conservation in \NCCA.

  {\bf State conservation is local: }
  If there is a state 
  $\alpha$ in a configuration,
  then there must be a state $\alpha$ close to it in the image. 
  Suppose this is not true. Then
  there are configurations where no $\alpha$ appears in the image of 
  a window of length $L$
  around the original $\alpha$, for arbitrarily large $L$. We may 
  choose $L$ large enough to
  assure that there is a word of length $n$ which is repeated 
  in the window; we cut
  the configuration between the repetitions (keeping one of them),
  and obtain a periodic 
  configuration {\em without} a state
  $\alpha$ in its image, which is a contradiction.

  {\bf The right PA for a state-conserving CA: }
  If state-conservation is local, then the most reasonable 
  way to look at the
  rule in terms of particles is to consider each state as a 
  single particle, with
  different states corresponding to different types of 
  particles (and,
  in fact, the numerical values of the states turn out to be 
  irrelevant, and could
  be replaced by colors or letters). But then we would like to 
  have a kind of particle automaton
  that takes a particle of type $\alpha$ and a surrounding 
  configuration, and moves it to the
  position of a particle of type $\alpha$ in the image; it would 
  be defined by a tuple 
  $(q,\{-\ell,\sdot,r\},g)$, with $g:Q^\ell\times Q^r \to N$, and $g$ would 
  determine the new position of
  the particle currently located at the origin in $N$.
  If we want to keep the image of the states as occupancy 
  numbers, then what we 
  would want is a particle automaton that moves all the 
  particles of a cell 
  together: instead of evaluating to an arbitrary vector
  in $N^i$, each $g_i$ (in the notation for PA) would evaluate to 
  a vector $(\underbrace{j,j,\sdot,j}_i)$, for some $j\in N$.

  {\bf Constructing the right PA: }
  Unfortunately, this is not the PA that will result from the 
  application of Theorem
  \ref{thm:pa4ca} (unless we have $q=2$, or a particular case 
  like the shift). However,
  the construction given in the proof of Theorem 
  \ref{thm:pa4ca} may be fixed for
  the case of state-conserving CA: for each configuration, 
  we assign now to each
  particle (\ie, to each state in each cell) the position of its 
  correlative particle (from left to right)
  in the image of the configuration (this can be done, thanks 
  to the 
  state-conservation). The proof then proceeds exactly
  in the same way, and the resulting PA will have the same neighborhood
  (and the particles will arrive in the same zone inside it) as stated
  in Corollary \ref{cor:regular}. 
  In general, order will not be preserved.

  \stepcounter{example}
  {\bf Example \arabic{example}: } 
  Consider $f\in$ \CAa{3}{3} with code \#6768185473053, and offset 1.
  Using Theorem \ref{thm:pa4ca} we obtain the motion 
  representation
  \[
   M_{14} \;=\;
	\{
	 \overset{\overset{1}{\curvearrowright}}{21}
	\}
  \]
  However, this CA is state-conserving: a `2' will travel to the 
  right as long as it is immersed in a background of 1's. 
  Depending on the application, it may
  be therefore more appropriate to use the special version 
  of the construction (as described above), and obtain the motion
  representation
  \[
   M_{15} \;=\;
	 \{
     \overset{\overset{2}{\curvearrowright}}{21},
	  \overset{\overset{ }{\curvearrowleft}}{21}
	 \}
  \]
  where the numbers in the arrows might be dropped, since they will 
  always represent
  the motion of the complete ``particle'', 1 or 2.
  \qed

  \subsection{Momentum Conservation}

  So far we have considered one additive conserved quantity, the mass.
  It is natural to ask about other quantities that frequently follow
  conservation laws, as, for instance, momentum. Notice that this question does 
  not apply to CA, but is natural for a PA. 
  
  In a PA $G=(q,N,(g_i)_{i=1,\sdot,q-1})$ with $N=\{-\ell,\sdot,r\}$, 
  we define the velocity of a particle at a given time step as the 
  difference between its position at the next time step 
  and its current position; equivalently, as the value that $g_i$
  assigns to it. Thus, the sum of the velocities of the particles 
  at a cell $i$ of a configuration $c \in S^\Zset$ is 
  \[
   V_i(c) \;=\; \sum_{k=1}^{c_i} [g_{c_i}(c_{i-\ell},\sdot,c_{i-1},c_{i+1},\sdot,c_{i+r})]_k
  \]
  We will say that a PA preserves the momentum if, and only if,
  \begin{equation}
    \sum_{i=0}^{p(c)-1} V_i(c)
	 \,=\,
	 \sum_{i=0}^{p(c)-1} V_i(c')
     \quad \forall c\in C_P,
	 \label{eq:defmom}
  \end{equation}
  where $C_P$ are the periodic configurations of $Q^\Zset$ and
  $c' = G(c)$.  
  In other words, the function $\phi:Q^{\ell+r+1}\to \Zset$ defined by
  \[
   \phi(x_0,\sdot,x_{\ell+r}) \;=\; 
	 \sum_{k=1}^{x_\ell} [g_{x_l}(x_0,\sdot,x_{\ell-1},x_{\ell+1},\sdot,x_{\ell+r})]_k
  \]
  is asked to be the density of an additive preserved quantity for $\Pi(G)$.
  Direct application of Theorem \ref{teo:hattori} yields the
  following proposition.
  \begin{prop}
   Momentum preservation is a decidable property of PA.
  \end{prop}

  In fact, in spite of being defined in terms of the particle representation,
  momentum preservation depends only on the conservative CA we are representing, 
  as shown by the next theorem. Notice that this is not true for non-conservative CA.

  \begin{thm}
  Let $F$ be a conservative CA and let $G$ and $G'$ be PA such that 
  $\Pi(G)=\Pi(G')=F$. Then the following three
  are equivalent:
   \renewcommand{\labelenumi}{(\roman{enumi})}
   \begin{enumerate}
	 \item
     $G$ preserves momentum
	 \item
	  $G'$ preserves momentum
	 \item
	  $F$ verifies
	  \[
		 \sum_{i\in\Zset} \sum_{j\leq i} c_j-c'_j = \sum_{i\in\Zset} \sum_{j\leq i} c'_j-c''_j
		 \qquad \forall c\in C_F \textrm{, where } c'=F(c),\, c''=F^2(c)
		\]
	\end{enumerate}
  \end{thm}
  \begin{pf*}{Proof.}
  Since momentum is an additive quantity, Theorem \ref{teo:fin} implies
  that condition (\ref{eq:defmom}) can be tested on the configurations
  of $C_F$ instead of $C_P$. Therefore, (i)$\Leftrightarrow$(ii) follows from
  (i)$\Leftrightarrow$(iii) together with the fact that $\Pi(G)=\Pi(G')$, and
  we just need to show (i)$\Leftrightarrow$(iii). 
  
  If $F=\Pi(G)$, then $F(c)=G(c)$ for all $c\in Q^\Zset$, and we can forget
  $F$. Let $c$ be a finite configuration, $c'=G(c)$ and $c''=G(c')$.
  Then what we must show is that 
	  \[
	    \sum_{i\in\Zset} V_i(c)
		 \,=\,
	    \sum_{i\in\Zset} V_i(c')
		 \, \iff \, 
		 \sum_{i\in\Zset} \sum_{j\leq i} c_j-c'_j = \sum_{i\in\Zset} \sum_{j\leq i} c'_j-c''_j
	  \]
  In fact, what we have is that
   \begin{equation}
	\sum_{i\in\Zset} \sum_{j\leq i} c_j-c'_j 
	\;=\;
	\sum_{i\in\Zset} V_i(c)
	\;=\;
	\sum_{i\in\Zset} \sum_{j=1}^{c_i} [g_{c_i}(c_{i-\ell},\sdot,c_{i-1},c_{i+1},\sdot,c_{i+r})]_j
	 \label{eq:demomom}
	\end{equation}
	To see this, consider the contribution of each particle in $c$ to each of
	the sums. On the left side, the $j$-th particle at $c_k$ contributes 
	its displacement, $v=[g_{c_k}(c_{k-\ell},\sdot,c_{k-1},c_{k+1},\sdot,c_{k+r})]_j$. Without loss
	of generality, suppose $v\geq 0$. The left side of (\ref{eq:demomom})
	is the addition, over $i\in \Zset$, of $\sum_{j\leq i} c_j-c'_j$, the 
	{\em accumulated difference} between $c$ and $c'$. The particle is moving 
	from $k$ to $k+v$; thus, it contributes $+1$ to this sum, for the
	$v$ terms corresponding to $i=k,\sdot,k+v-1$, \ie, its movement
	contributes with $v$ to the total sum.
  \qed\end{pf*}

  \begin{cor}
  Momentum preservation is a decidable property in \NCCA.
  \end{cor}

  \stepcounter{example}
  {\bf Example \arabic{example}: } 
  Most of the momentum preserving NCPA with small neighborhoods
  are trivial (identity, shifts); the rest of the cases consist of rules that
  only allow movements with zero sum, as in the following motion representations.
  \[
   M_{16} \;=\;
	\{
	 \mbox{\jumponeleft{00110}{5}{2}}, \mbox{\jumponeright{00110}{5}{3}}
	\}
	\qquad,\qquad
   M_{17} \;=\;
	\{
	  \overset{\overset{ }{\curvearrowleft\curvearrowright}}{120}
	\}
  \]
  Of course, more sophisticated CA with momentum preservation
  can be constructed for larger neighborhoods.
  \qed

\section{The Monotone Case}
  \label{sec:dec}

  In this section we deal with the characterization and particle representation
  of monotone one-dimensional CA. In fact, we will talk almost exclusively about
  non-increasing CA (\DCA), but it must be noticed that this
  is equivalent to talking about non-decreasing CA (or, at least, each result
  about the former translates into a result about the latter). In fact,
  there is a one-to-one correspondence between the elements of 
  \DCA and those of \ICA, by replacing ``particles'' with ``non-particles''
  and vice-versa: for each
  $f\in$ \DCA we have a $\tilde{f}\in$ \ICA defined by
  \[
  \tilde{f}(x_1,\sdot,x_n) = q-1 \,-\, f( q-1-x_1,\sdot,q-1-x_n )
  \].

 \subsection{An Only Sufficient Condition}

  A first idea for a characterization of monotone CA would be the
  replacement of the equality for an inequality in condition (\ref{eq:ncsc}),
  \ie, we would like to say that $f$ is in \DCAa{q}{n}
  if and only if, for all $(x_1,\sdot,x_n)\in Q^n$,
   \begin{equation}
    f(x_1,\sdot,x_n) \;\leq\; x_1+\sum_{k=1}^{n-1} 
    \left\{
      f(\underbrace{0,\sdot,0}_{n-k},x_2,\sdot,x_{k+1})
      - f(\underbrace{0,\sdot,0}_{n-k},x_1,\sdot,x_k)
     \right\}
     \label{eq:nni}
    \end{equation}
	However, this condition is only sufficient. To see that it is sufficient,
	consider any periodic configuration, and add both sides of the inequality 
	along a whole period: all the terms in the sums of the right side will
	cancel, and we obtain (\ref{eq:naq}). On the other hand, the condition
	is not necessary, as shown by the following example.

  \stepcounter{example}
  {\bf Example \arabic{example}: } 
  Let $F$ be the elementary CA 72. Here $q=2$, the offset is 1, and
  \[
    f(x_1,x_2,x_3) \;=\; \left\{  
	  \begin{array}{cl} 1 & \textrm{ if } x_2=1 \textrm{ and }x_1+x_3=1 \\
	                    0 & \textrm{ otherwise}
	\end{array} \right.
  \]
  It is easy to see that $f\in$ \DCA: the only way in which
  a cell can get a 1 in the image is to already have one in the preimage.
  On the other hand, condition (\ref{eq:nni}) is not verified:
  \[
   1 \;=\; f(1,1,0) \;>\; 1 + f(0,1,0)+f(0,0,1)-f(0,1,1)-f(0,0,1) \;=\; 0
  \]\qed

 \subsection{Domination and maximal elements in \DCA}

  We will say that $h\in$ \CAa{q}{n} {\em dominates}
  $f\in$ \CAa{q}{n} if $f(w)\leq h(w)$ for all $w\in Q^n$.
  It is easy to see that any rule $f\in$ \CA
  which is dominated by a rule $h\in$ \NCCA
  belongs to \DCA. This arises a natural question: 
  are all elements of \DCAa{q}{n} dominated by
  elements of \NCCAa{q}{n}? In fact, this is true for
  $q=2$, $n=3$ (the elementary CA). However, the general 
  answer is negative:

  \stepcounter{example}
  {\bf Example \arabic{example}: }
  Consider $f\in$ \CAa{3}{2} defined by 
  \[
   f(x_1,x_2) = \left\{ \begin{array}{cl}
	  2 & \textrm{ if } x_1=2 \\
	  1 & \textrm{ if } x_1\in \{0,1\} \textrm{ and } x_2=1 \\
	  0 & \textrm{ otherwise}
	  \end{array}
	  \right.
  \]
  It is easy to see that $f\in$ \DCA, since
  $f(x_1,x_2)\leq x_1$. 
  Now suppose that there is $h\in$ \NCCAa{3}{2} such that $f\leq h$.
  Since $h$ is conservative, it must verify $h(0,0)=0$, $h(1,1)=1$ and
  $h(2,2)=2$; since it dominates $f$, it must also verify $h(2,0)=2$ and
  $h(2,1)=2$. From (\ref{eq:ncsc}) we have
  \[
   \begin{array}{l}
   2=h(2,0)=2+h(0,0)-h(0,2) \quad \textrm{, and } \quad 
	\\
	2=h(2,1)=2+h(0,1)-h(0,2)
	\end{array}
  \]
  Since $h(0,0)=0$, we get $h(0,2)=0$ and hence $h(0,1)=0$, which is
  less than $f(0,1)=1$.

  The CA defined by $f$ (with offset 0) keeps the 2's 
  untouched, while the 1's travel to the left until they hit a 2, and 
  disappear (it is the projection of the PA described by $M_5$ in
  Section \ref{sec:defpa}). Most small examples of non-dominated 
  rules in \DCA follow
  the same pattern: a wall of some kind, and particles that
  move until they meet it and disappear. Intuitively, a CA dominating
  them would have to preserve everything which is being destroyed 
  (and it cannot do this only ``next to the wall''), at the same time 
  that it follows 
  the particles in 
  their movement; therefore, it would have to increase the total mass.
  \qed

  We will say that $f\in$ \DCAa{q}{n} is {\em maximal} if
  it is not dominated by another element of \DCAa{q}{n}.

 \subsection{A Characterization of \DCA}
  In the following proposition we show that monotony is
  decidable. Unfortunately, our characterization is computationally
  useful only for small values of $q$ and $n$.

  \begin{thm}
   Let $f$ be a local rule in \CAa{q}{n}. Then $f\in$ \DCAa{q}{n} if and only if
  \begin{equation}
   \sum_{k=0}^{|w|-1} f(w/w)_k  \; \leq \; \sum_{k=0}^{|w|-1} w_k,
     \qquad \forall w\in L(q,n)
 	  \label{eq:cdca}
  \end{equation}
  where $L(q,n)$ is the set of all words in $Q^*$ such that
  \[
   \left(w_{i \textrm{ mod } |w|},\sdot, w_{(i+n-2) \textrm{ mod } |w|} \right)
	\neq 
	\left(w_{j \textrm{ mod } |w|},\sdot, w_{(j+n-2) \textrm{ mod } |w|} \right)
	\quad \textrm{ for } i\neq j
  \]
  \label{thm:carac}
  \end{thm}
  In other words, $L(q,n)$ are the words that do not repeat
  a subword of length $n-1$, when considered as circular words. They
  correspond to all the cycles in the de Bruijn graph $B(n-2,q)$ that do not
  repeat edges; their maximum length is that of the Eulerian paths
  in $B(n-2,q)$, which is $q^{n-1}$.

  {\bf Proof. }
  Since (\ref{eq:cdca}) is the restriction of (\ref{eq:defdec}) to some
  configurations, the condition is obviously necessary. To show its
  sufficiency, we have to consider a word $w\in S^*\setminus L(q,n)$.
  Then $w$ must have a subword $\alpha$ of length $n-1$ which occurs twice 
  in $w$. There are two cases: the occurrences of $\alpha$ overlap, or
  they do not.
  
  {\bf Case A: No overlap. }
  Without loss of generality, we can assume that $w$ begins with $\alpha$,
  and write it as $w=\alpha u \alpha v$. In addition, we can assume that
  both $\alpha u\,\in L(q,n)$ and $\alpha v\,\in L(q,n)$: if not, we apply 
  the whole argument to them (recursively). Thus, they verify (\ref{eq:cdca}),
  and we have
  \begin{eqnarray*}
   \sum_{k=0}^{|w|-1} f(w/w)_k  & 
	\;=\;  &
	\sum_{k=0}^{|w|-1} (\alpha u \alpha v / \alpha u \alpha v )_k 
	\\
	&
	\;=\; &
	\sum_{k=0}^{|\alpha u|-1} (\alpha u / \alpha )_k + 
	\sum_{k=0}^{|\alpha v|-1} (\alpha v / \alpha )_k 
	\\
	&
	 \;\leq \; &
	\sum_{k=0}^{|\alpha u|-1} (\alpha u )_k + 
	\sum_{k=0}^{|\alpha v|-1} (\alpha v )_k 
	 \;=\; 
   \sum_{k=0}^{|w|-1} w_k
  \end{eqnarray*}

  {\bf Case B: With overlap. }
  In this case, two occurrences of $\alpha$ overlap. It follows that $\alpha$ has 
  a prefix $\beta$ such that $\alpha_{i} = \beta_{i \textrm{ mod } |\beta|}$. Note that
  \[
   f(w/w) \;=\;
	f(w/\alpha) \;=\;
	f(\beta \alpha u / \alpha ) \;=\;
	f(\beta / \alpha) f(\alpha u / \alpha )
  \] 
  As before, we can assume that inequality (\ref{eq:cdca}) is satisfied for
  the words $\alpha u$ and $\beta$ (in fact, since $|\beta|<n-1$, we apply
  it to $\beta^m$, with $m|\beta|\geq n-1$, and divide by $m$ to obtain 
  it for $\beta$). Thus we get
  \begin{eqnarray*}
   \sum_{k=0}^{|w|-1} f(w/w)_k &
	\;=\;
	\sum_{k=0}^{|\beta|-1} (\beta / \alpha )_k + 
	\sum_{k=0}^{|\alpha u|-1} (\alpha u / \alpha )_k 
 \\
  & 
	 \;\leq \; 
	\sum_{k=0}^{|\beta|-1} \beta_k + 
	\sum_{k=0}^{|\alpha u|-1} (\alpha u )_k 
	&
	 \;=\; 
   \sum_{k=0}^{|w|-1} w_k
\\
\end{eqnarray*}
  \qed

  As we said before, this test is not very practical, since it involves checking
  the condition for a large number of words, which grows as $q^{q^n}$. A careful
  listing of the cycles in de Bruijn graph $B(n-2,q)$ could make this a bit 
  lower, but not much: there are at least $(q!)^{q^{n-3}}q^{-(n-2)}$ Eulerian cycles in $B(n-2,q)$. 
  
 \subsection{Particle Representation for \DCA}
  \label{subs:pa4dca}

 Once we have characterized one-dimensional monotone CA, and since we know 
 that one-dimensional conservative CA can be represented through particle
 automata, a natural problem to consider is the representation of
 non-increasing CA in terms of the movement of particles.
 Clearly, the projection of any PA is a non-increasing CA. For the converse,
 we will show how to construct a PA for a given non-increasing CA.

 More precisely, we will show how to associate to any cell in the image of a
 configuration, the location that its particles had in that configuration (this
 is a particular case of the {\em particle identifications} defined in \cite{kurka2}). 
 Consider a word $w\in Q^n$, with image $f(w)>0$. Since the CA is non-increasing,
 we have that $f(w) \leq \sum_{i=0}^{n-1} w_i$. We will impose that the $f(w)$
 particles in the image cell must come from the cells in which $w_0,\sdot,w_{n-1}$
 were located. Moreover, the particles will be assumed to be contiguous in the preimage.
 Then their location in $w$ is defined by a single value $s(w)$, 
 $s(w)\in\{0,\sdot,\sum_{i=0}^{n-1} w_i-f(w)\}$, as shown in the following
 scheme:
\begin{equation}
 \underbrace{ 
  \overbrace{\bullet \sdot \bullet}^{w_0}  \,
  \overbrace{\bullet \sdot \bullet}^{w_1}  \,
  \sdot \bullet}_{s(w)}
  \;
 \underbrace{\bullet \sdot \bullet}_{f(w)}
 \;
 \sdot
  \overbrace{\bullet \sdot \bullet}^{w_{n-1}} 
 \label{eq:scheme}
\end{equation}
Denote by $p(w,i)$ the number of particles contributed by each $w_i$ to
the cell that holds $f(w)$ in the image; this can be better understood
in the scheme above, by seeing how many of the $f(w)$ selected particles
fall in $i$, but its precise value can be written as
\begin{equation}
 p(w,i) \;=\; \left( 
   \textrm{min }\{ s(w)+f(w), \sum_{j\leq i} w_j \}
	\quad - \quad
	\textrm{max }\{ s(w), \sum_{j < i} w_j \}
	\right)_+
 \label{eq:pwi}
\end{equation}
and we have $f(w) = \sum_{i=0}^{n-1} p(w,i)$.

\begin{thm}
 Let $f\in$ \CAa{q}{n}. Then a 
 necessary and sufficient condition for $f$ to be in \DCA
 is that there exists $s:Q^n\to \Zset$, with $s(w)\in\{0,\sdot,
 \sum_{i=0}^{n-1} w_i - f(w)\}$ for all $w=w_0,\sdot,w_{n-1}$ and
 $s(w)=0$ for all $w$ with $f(w)=0$, such that
 $p(w,i)$ defined by (\ref{eq:pwi}) verifies
 \begin{equation}
  \sum_{i=0}^{n-1} p((w_{n-1-i},\sdot,w_{2n-2-i}), i) \;\leq \; w_{n-1}
  \qquad \textrm{ for all } w_0,\sdot,w_{2n-2} \in Q
  \label{eq:condthm}
 \end{equation}
 \label{thm:eqdca}
\end{thm}
 
 {\bf Proof. }

 {\bf Sufficient condition. }Suppose that there exists $s$ with
 these properties, and consider any finite configuration $c\in Q^\Zset$.
 Then
  \begin{eqnarray*}
     \sum_{j\in \Zset} f(c_j,\sdot,c_{j+n-1})  &
  \;=\;
  \sum_{j\in \Zset} \sum_{i=0}^{n-1} p( (c_j,\sdot,c_{j+n-1}), i) 
 \\
  & 
  \;=\;
  \sum_{j\in \Zset} \sum_{i=0}^{n-1} p( (c_{j-i},\sdot,c_{j-i+n-1}), i) 
	&
  \;\leq\;
  \sum_{j\in \Zset} c_j
\\
\end{eqnarray*}

 {\bf Necessary condition. }Suppose that $f\in$ \DCA. For
 each $w=w_0,\sdot,w_{n-1}$ in $Q^n$ define $s(w)=0$ if $f(w)=0$,
 and
 \begin{equation}
	s(w) \;=\; 
	 \sum_{i=0}^{n-1} w_i +
       \min_{v\in Q^*} \left\{
      \sum_{i=0}^{|v|-1} v_i
	  - \sum_{i=0}^{n+|v|-1} f(wv/ 0^{n-1})_i
           \right\}
	 \label{eq:defsw}
 \end{equation}
 otherwise. It verifies
 \[
  0 
  \;\leq\;
  s(w) 
  \;\leq\;
    \sum_{i=0}^{n-1} w_i
	  - \sum_{i=0}^{n-1} f(w/ 0^{n-1})_i
  \;\leq\;
    \sum_{i=0}^{n-1} w_i - f(w)
 \]
 where the lower bound follows from $f\in$ \DCA, and
 the upper one is obtained by taking $v$ as the empty word (or
 as ``0'') in the definition of $s(w)$.
 
 It is important to notice that $s$ can be determined with
 a finite computation: if $v$ is a word of minimal length where
 the minimum is reached, then $|v|\leq q^{n-1}$. The situation
 is similar to the proof of Theorem \ref{thm:carac}; since we are not 
 taking circular words, we need a word of length greater $q^{n-1}+n-1$
 to assure the existence of a repeated subword $\alpha$ of
 length $n-1$, and this length is reached by $wv$. Then
 $wv$ can be written as $x\alpha y \alpha z$, and we can
 define $v'$ such that $wv'=x\alpha z$, which verifies
 \[
  \begin{array}{l}
   \displaystyle
	 \sum_{i=0}^{n-1} w_i + \sum_{i=0}^{|v'|-1} v'_i
	  - \sum_{i=0}^{n+|v'|-1} f(wv'/ 0^{n-1})_i
   \\
   \displaystyle
	 \textrm{ }\qquad =\;
	 s(w) + \sum_{i=0}^{|\alpha y|-1} (\alpha y)_i 
	     - \sum_{i=0}^{|\alpha y|-1} f(\alpha y / \alpha )_i
	\;\leq\; 
   s(w)
  \end{array}
 \]
 where the last inequality is implied by the non-decreasing
 property: if it is false, then the periodic configuration
 with periodic pattern $\alpha y$ contradicts the monotony. Here we
 assumed that the two occurrences of $\alpha$ do not overlap;
 if they do, the reasoning is similar, akin to that in case
 B of Theorem \ref{thm:carac}. In both cases, what we
 get is a violation of the minimality of $|v|$.

 We still have to show that $s(w)$ verifies (\ref{eq:condthm}).
 This follows from the fact that different particles in the image
 take different particles as their preimages. Consider a configuration
 $c\in Q^\Zset$ and two positive coordinates in $F(c)$; we may 
 assume that the offset is 0 (\ie, $F(c)_0=f(c_0,\sdot,c_{n-1})$),
 and that the two positions are 0 and some $k>0$. Consider a scheme
 like (\ref{eq:scheme}), with the $\sum_{i=0}^{k+n-1} c_i$ particles
 of the positions $0,\sdot,c_{k+n-1}$ aligned in a row, and number
 them from left to right. Then the $f(c_0,\sdot,c_{n-1})$ particles
 landing at position $0$ are taken as the particles labeled
 with numbers 
 \[
  s(c_0,\sdot,c_{n-1})+1,\sdot,s(c_0,\sdot,c_{n-1})+f(c_0,\sdot,c_{n-1})
 \]
 while the $f(c_k,\sdot,c_{k+n-1})$ particles landing at $k$
 are taken as the particles labeled with the numbers
 \[
  \sum_{i=0}^{k-1} c_k + s(c_k,\sdot,c_{k+n-1})+1, \sdot,
  \sum_{i=0}^{k-1} c_k + s(c_k,\sdot,c_{k+n-1})+f(c_k,\sdot,c_{k+n-1})
 \]
All we need to show is that they do not overlap, \ie, 
\[
 s(c_0,\sdot,c_{n-1})+f(c_0,\sdot,c_{n-1})
 \; \leq \;
 \sum_{i=0}^{k-1} c_k + s(c_k,\sdot,c_{k+n-1})
\]
Consider any word $v\in Q^n$, and put the word 
$v'=(c_n,\sdot,c_{k+n-1},v_0,\sdot,v_{|v|-1})$
in (\ref{eq:defsw}) for the definition of $s(c_0,\sdot,c_{n-1})$.
We obtain that
 \[
  \begin{array}{lcl}
  \displaystyle
 s(c_0,\sdot,c_{n-1}) 
	 & \leq & 
	   \displaystyle
	 \sum_{i=0}^{n-1} c_i + \sum_{i=0}^{|v'|-1} v'_i
	  - \sum_{i=0}^{n+|v'|-1} f(c_0,\sdot,c_{n-1},v'/ 0^{n-1})_i
    \\
	 & = & 
	   \displaystyle
    \sum_{i=0}^{k-1} c_i + \sum_{i=k}^{k+n-1} c_i + \sum_{i=0}^{|v|-1} v_i
	 - \sum_{i=0}^{k-1} f(c_i,\sdot,c_{i+n-1}) 
	 \\
	 & &
	   \displaystyle
 \textrm{ }\qquad  
    - \sum_{i=0}^{|v|+n-1} f(c_k,\sdot,c_{k+n-1},v/ 0^{n-1})_i
  \end{array}
 \]
 Thus, any $v\in Q^*$ verifies
 \[
  \begin{array}{l}
   \displaystyle
  \sum_{i=k}^{k+n-1} c_i + \sum_{i=0}^{|v|-1} v_i - \sum_{i=0}^{|v|+n-1} 
	 f(c_k,\sdot,c_{k+n-1},v/ 0^{n-1})_i
	 \\
	 \textrm{ }\qquad 
   \displaystyle
	 \;\geq\; s(c_0,\sdot,c_{n-1}) - \sum_{i=0}^{k-1} c_i + \sum_{i=0}^{k-1} f(c_i,\sdot,c_{i+n-1})
	 \end{array}
 \]
 and therefore, taking the minimum over all $v$,
 \[
  \begin{array}{l}
   \displaystyle
  s(c_k,\sdot,c_{k+n-1}) 
  \\
   \textrm{ }\qquad 
	\displaystyle
	\geq\; 
  s(c_0,\sdot,c_{n-1}) - \sum_{i=0}^{k-1} c_i + 
  \sum_{i=0}^{k-1} f(c_i,\sdot,c_{i+n-1})
    \\
   \textrm{ }\qquad 
	\displaystyle
  \geq\; 
  s(c_0,\sdot,c_{n-1}) - \sum_{i=0}^{k-1} c_i + f(c_0,\sdot,c_{n-1})
  \end{array}
 \]
 \qed
 
\begin{cor}
 Let $F$ be a CA with local rule $f\in$ \DCAa{q}{n}. Then 
 there exists $G\in$ \PAa{q}{2n-1} such that $\Pi(G)=F$.
\end{cor}

  {\bf Proof. }
  The functions $s$ and $p$ as in Theorem \ref{thm:eqdca} determine the
  particle automaton. For any configuration 
  $w_0,\sdot,w_{n-1},\sdot,w_{2n-2}$ with $w_{n-1}>0$, we
  define the motion of the $w_{n-1}$ particles as follows:
  the number of particles moving from $n-1$ to $i$, $i=0,\sdot,n-1$,
  is given by $p_i=p( (w_{n-1-i},\sdot,w_{2n-2-i}), i)$, and the 
  number of particles that die (move to $\dagger$) is given by
  $
   w_{n-1} \;-\; \sum_{i=0}^{n-1} p_i
  $. In other words, we define a $g_{w_{n-1}}$ such that
  \[
   \begin{array}{cl}
    &
    \displaystyle
    \#\{ k: [g_{w_{n-1}}(w_0,\sdot,w_{n-2},w_{n},\sdot,w_{2n-2})]_k=i\} = p_i
    \\
    \textrm{and }
    &
    \displaystyle
    \#\{ k: [g_{w_{n-1}}(w_0,\sdot,w_{n-2},w_{n},\sdot,w_{2n-2})]_k=i\} = w_{n-1} \;-\; \sum_{i=0}^{n-1} p_i
    \end{array}
  \]
  \qed
 
 \begin{cor}[Another test for CA monotony] The previous theorems
 establish the equivalence between (one-dimensional) 
 non-increasing CA and particle automata. This provides
 an alternative characterization of \DCA: to see
 if a certain CA belongs to this class, check the existence of
 a function $s$ with the properties of Theorem \ref{thm:eqdca}.
 Finding it in the way described in the proof would need (in 
 definition \ref{eq:defsw}) as much work as the characterization
 in Theorem \ref{thm:carac}; it is usually easier to check for its existence
 by testing the possible functions $s$. Since $s:Q^n\to\Zset$,
 and $0\leq s(w)\leq \sum w_i - f(w)$, an upper bound for the number
 of possible $s$ is $(qn)^{q^n}$, which is more than the work
 in Theorem \ref{thm:carac}, but using that $s(w)=0$, $p(w,i)=0$
 for $f(w)=0$, and applying the necessary conditions of Theorem
 \ref{thm:eqdca}, the possibilites for $s$ can be drastically reduced
 in a practical implementation.
 \end{cor}

 {\bf Remark: There is no canonical PA. } As seen
 in Section \ref{sec:ncca}, we
 can always take a ``canonical'' particle representation for a
 (one-dimensional) conservative CA,
 which is unique, and is characterized by the preservation of the order
 of the particles. In the monotone case, there is no canonical form: 
  for instance, when
 two particles are close to each other and one disappears, there is
 an arbitrary decision favouring the survival of one of them.

 \begin{table}[htb]
  \begin{center}
   \begin{tabular}{||c|c||c|c|c|c||} \hline \hline
	 \label{tab:demog}
	  $q$ & $n$ & \NCCA & state-cons. CA & \DCA & maximal \DCA \\
	  \hline \hline
	  2 & 2 &       2 &      2 &       6 &  2 \\
	  2 & 3 &       5 &      5 &      46 &  5 \\
	  2 & 4 &      22 &     22 &    2756 &  38 \\
	  2 & 5 &     428 &    428 &       ? &  ? \\
	  2 & 6 &  133184 & 133184 &       ? &  ? \\
	  \hline
	  3 & 2 &       4 &      2 &     708 &  6 \\
	  3 & 3 &     144 &     15 &       ? &  ? \\
	  3 & 4 & 5448642 &      ? &       ? &  ? \\
	  \hline
	  4 & 2 &      10 &      2 & 3732576 &  ? \\
	  4 & 3 &   89588 &     89 &       ? &  ? \\
	  \hline \hline
	 \end{tabular}
  \caption{Some demographics: Number of one-dimensional conservative, state-conserving, 
  non-increasing, and maximal non-increasing rules, for some $q$ and $n$.}
  \end{center}
 \end{table}

 \section{Conservation Associated to Blocks}
 \label{sec:block}

  In \cite{moreiratcs}, Theorem 6, a method was described to 
  decide, for a given CA, whether its states can be relabelled 
  with integer numbers so as to make the CA number-conserving;
  in Section \ref{sec:ncca}, we see that this can be used, in 
  turn, to see if the dynamics of
  the original CA can be understood in terms of an operator 
  acting on a system of indestructible particles.

  In this section we consider a generalization of that theorem, giving
  values to the words of a given length, instead of the single states; we
  want this function of the blocks to be preserved by the iteration of the CA.
  In other words, we are looking for an additive conserved quantity; however, we
  will impose a further condition: it must separate the images of the blocks
  with different values at the origin. 
  
  We want to view the function as an
  assignment of a value to the state in a cell, but a value which
  depends on the states of its neighbors: the same 
  state $a$ may represent 2 or 3 particles, depending on whether or not it 
  is followed by, say, another state $a$. 
  Thus we can detect if the given CA may be seen as the projection, to 
  fewer states, of a conservative CA which in turn is seen as an 
  interaction of particles, restricted to certain configurations. 
  In particular, this may automatically detect particle-like behavior in
  some non-conservative CA.

  \begin{thm}
  Let $F$ be a one dimensional CA with state set $S\subset \Zset$, 
  and let $b$ be a positive integer. 
  Then it may be decided whether or not there is a 
  $b$-neighborhood-dependent relabelling of the states of $S$, which 
  distinguishes the states, and whose sum is preserved by $F$.
   \label{teo:block}
  \end{thm}

  {\bf Proof. }
  Consider a cellular automaton $F$ with states $S\subset \Zset$, 
  neighborhood size $n$, and local rule 
  $f\in$ \CAa{q}{n}. Without loss of generality, the offset
  can be assumed to be 0. Let $\phi$ be a relabelling map,
  $\phi:S^b\to \Zset$.
  For $c\in S^{\Zset}$, we define $\Phi(c)$ as $\Phi(c)_i=\phi(c_i,\sdot,c_{i+b-1})$.
  Let $F_\phi$ be the induced CA that makes the diagram
  \[
   \begin{array}{ccc}
	 S^\Zset & \stackrel{\Phi}{\rightarrow} & \Phi(S^\Zset) \\
	 F \downarrow & & F_\phi \downarrow \\
	 S^\Zset & \stackrel{\Phi}{\rightarrow} & \Phi(S^\Zset)
	\end{array}
  \]
  commute. $F_\phi$ acts on the subshift $\Phi(S^\Zset)$, and has the same
  neighborhood as $F$.
  We want to determine possible mappings $\phi$ that would make $F_\phi$
  number-conserving. In other words, we want to find $\phi$ such that
  \begin{equation}
   \sum_{k=0}^{p(c)-1} \phi(c_i,\sdot,c_{i+b-1})
	\;=\;
   \sum_{k=0}^{p(c)-1} \phi(f(c_i\sdot c_{i+n-1}),\sdot,f(c_{i+b-1}\sdot c_{i+b+n-2}))
	\label{eq:pe}
  \end{equation}
  for all $c\in C_P$.
  Fix an arbitrary $\alpha\in S$. From Theorem \ref{teo:hattori} we have that
  a necessary and sufficient condition for (\ref{eq:pe}) is that
  \begin{equation}
  \begin{array}{l}
   \displaystyle
   \phi_f(a_0,\sdot,a_{b+n-2})
	-\phi(a_0,\sdot,a_{b-1}) = \\
   \displaystyle
	 \sum_{i=1}^{n+b-2} \big\{ -\phi_f(\underbrace{\alpha,\sdot,\alpha}_i,a_0,\sdot,a_{n+b-2-i})
	  + \phi_f(\underbrace{\alpha,\sdot,\alpha}_i,a_1,\sdot,a_{n+b-1-i}) 
	 \big\} \\
   \displaystyle
	 +\sum_{i=1}^{b-1} \phi(\underbrace{\alpha,\sdot,\alpha}_{i},a_0,\sdot,a_{b-i-1})
	                  -\phi(\underbrace{\alpha,\sdot,\alpha}_{i},a_1,\sdot,a_{b-i})
  \end{array}
   \label{eq:te}
  \end{equation}
  holds for all $a_0,\sdot,a_{n+b-2}\in S$; here we use the notation
  \[
   \phi_f(x_0,\sdot,x_{b+n-2})=\phi(f(x_0,\sdot,x_{n-1}),
   \sdot,f(x_{b-1},\sdot,x_{b+n-2})).
  \]
 
  This condition can be used to get an homogeneous linear
  system, where the $|S|^b$ unknown values are $\{\phi(w), w\in S^b\}$. 
  The solution will be a linear subspace $V\subset \Rset^{S^b}$; if
  $V=\{0\}$, then the algorithm gives a negative answer. If $V\neq \{0\}$,
  we are not ready yet: we demand the solution to distinguish between 
  different states in the first position
  (we are interpreting the image of the block $a_0,\sdot,a_{b-1}$ as the
  value assigned to $a_0$). Thus we ask that
  \[
     \phi\big(\{\alpha\}\times S^{b-1}\big) 
       \cap 
     \phi\big(\{\beta\}\times S^{b-1}\big) 
  \;=\; 
     \emptyset \qquad \forall \alpha\neq \beta.
  \]
  or, in other words, $\phi$ cannot belong to the hyperplane $\phi(u)=\phi(v)$,
  for $u,v\in S^b$, $u_0\neq v_0$. Let $\{E_{u,v}\}_{(u,v)\in I}$ be
  the collection of these hyperplanes, where $I=\{ (u,v):u,v\in S^b, u_0\neq v_0\}$.
  Then we want to see if $V\setminus ( \cup_{(u,v)\in I} E_{u,v} )\neq \emptyset$.
  Since each $V\cap E_{u,v}$ is a subspace of $V$, and linear spaces cannot be
  finite unions of proper subspaces, we have:
  \[
   \begin{array}{l}
	 \displaystyle
    V\setminus \big( \bigcup_{(u,v)\in I} E_{u,v} \big) = \emptyset
	 \iff
	 V = \bigcup_{(u,v)\in I} V\cap E_{u,v}
	 \\
	 \displaystyle
	 \iff
	 \exists (u,v)\in I : V = V\cap E_{u,v}
	 \iff
	 \exists (u,v)\in I : V \subseteq E_{u,v}
	\end{array}
  \]
  and this last condition can be checked by adding the hyperplane equations to
  the linear system.

  If solutions do exist, then there are integer solutions and they can
  be found in finite time. Any such solution will be the desired assignment 
  of values to the states in the original CA, that depend on their contexts,
  and whose sum is preserved by the dynamics of the CA.
  \qed 

  {\bf The range of values for $b$. }
  There is not always a value of $b$ that allows us to find a 
  number-conserving relabelling of the states: this is the case, for instance,
  of the trivial CA with states $\{0,1\}$, neighborhood $\{0\}$, and 
  $f(0)=f(1)=0$. A natural question is: For a given CA, how
  many values of $b$ must we check before we conclude that no $b$ will
  work? This is an open question at the moment of this writing, since we 
  do not currently have a bound
  for $b$; intuitively, we can expect that such a bound exists, since there is no 
  reason why the considered neighborhood should be arbitrarily large
  with respect to the neighborhood of the CA.
  The problem is akin to that stated (and left open) first in \cite{takesue} 
  and then in \cite{hattori} of bounding the range
  of the additive conserved quantities for the elementary reversible CA considered
  there; both are particular cases of the general problem of bounding the range of the
  additive conserved quantities of a given CA.

  {\bf Particle representation for the relabelled CA. }
  The construction given in Theorem \ref{thm:pa4ca}
  can be trivially restricted to any subset of the possible
  configurations of states; in particular, it can be
  restricted to the subshift $\Phi(S^\Zset)$, yielding a particle
  representation of the operation of our CA on the states
  and its neighbors.

\begin{figure}[htb]
  \begin{center}
   \resizebox{7cm}{!}{\includegraphics{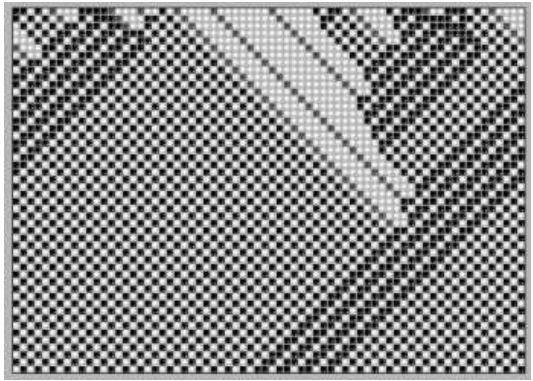}}
		\caption{Evolution of the rule from Example \ref{example:bloque}, with 
		  colors associated to 0, 1 and 2 particles.}
		\label{fig:ex3}
  \end{center}
\end{figure}

  \refstepcounter{example}
  \label{example:bloque}
  {\bf Example \arabic{example}: }
  Consider $f\in$ \CAa{3}{4} with code 64056. It 
  is not conservative, and it can be checked that there is no
  reassignment of values to its states that makes it number-conserving.
  However, there are solutions $\phi$ of the system given by equation (\ref{eq:te}), 
  for $b=3$. The smallest such $\phi$ (in norm) is
   \[
	 \begin{array}{rcl}
     \phi(0**)&=&0 \\
	  \phi(100) &=& 2 \\
	  \phi(101)=\phi(110)=\phi(111)&=&1
	 \end{array}
   \]
  We obtain for it the motion rule
  \[
   \overset{2}{\mbox{\jumponeright{2000}{4}{0}}} \quad
	\overset{\curvearrowright}{\jumponeright{200}{3}{0}}1 \quad
	\overset{\curvearrowright}{\jumponeright{200}{3}{0}}2 \quad
	\jumponeright{10}{2}{0} \quad
	\jumponeright{111}{3}{1} \quad
	\jumponeright{112}{3}{1}
	\]
  restricted to the configurations where no 1 is followed by two zeros, and
  all 2 are. This is a representation of the original CA in terms of the interaction
  of indestructible particles, and highlights one of the propagating defects
  in the evolution of this CA.
   \qed

\section{Conclusion}

This paper presented {\em particle automata} as simple
systems of interacting particles, that give a complementary view on the
behavior of conservative and monotone cellular automata. They are defined 
in terms of a global operator induced by a local rule with some finite
neighborhood (in analogy to the CA formalism), and may be easily
extended to higher dimensions or different topologies, as we expect to do
in a future article.

The special class of conservative PA is equivalent to the class
of conservative CA; we discussed several properties that the
conservative PA may exhibit, some of which
depend only on their CA projections. These intrinsic properties
of the particle representation include {\em anticipation}, the existence
of global anticipatory cycles, and {\em momentum conservation}, in a
natural sense defined here. Theorem \ref{thm:final} shows that the intrinsic
presence of the first two is equivalent, and that this presence is
a decidable property.
Some properties that are not intrinsic
are the preservation of order and the existence of local 
cycles, both of which can be always avoided by taking the ``canonical''
PA given by the equivalence result. Momentum conservation is an 
interesting case, since it turns out to depend only of the 
conservative CA, in spite of being defined (and making 
sense) in terms of a PA.

The conservative PA given by the construction algorithm may fail to be the
most useful one for a given CA. This is the case with 
state-conserving CA; for them, the
natural particle representation is one where the different states
represent different particles (instead of occupancy numbers).
We gave a characterization of state-conserving CA, and explained
the construction of appropriate PA for them.

 We have extended the work on conservative CA in
 two directions: on one hand, we have considered CA that are not
 conservative, but nearly so, since we can see them as the projection
 of particular configurations of a conservative CA with more states;
 their dynamics can be then understood in terms of indestructible
 particles.

 On the other hand, we have considered monotone CA, and we have shown that
 they can be characterized, and, moreover, that they can also
 be represented by means of a particle system. This equivalence is
 less obvious than in the conservative case; one of the difficulties is
 that no ``canonical'' PA exists for monotone CA.

 One open problem was already mentioned in the previous section: in 
 Theorem \ref{teo:block}, to find a bound for the block size $b$ that needs 
 to be considered before the existence of a solution $\phi$ is discarded. 
 Another natural problem, which arises from the text, is the following: 
 to extend Theorem \ref{teo:block} to allow $\phi$ to be monotone. This would 
 be very helpful, since it would help to detect particle-like behavior 
 that includes decay or annihilation. Such an extension is already 
 possible with the current results, but it would be computationally hard; 
 the challenge is to make it in an efficient way.

 {\em Please notice that some additional information, like rule lists,
  as well as some software, is available at our website:}
\begin{center}
{\tt http://www.dim.uchile.cl/\verb'~'anmoreir/ncca/}
\end{center}
 {\em Other software (like the C++ routines used to determine and solve
the linear systems for listing the CA rules) is available, by request, from
A. Moreira.}

\section*{Acknowledgments}

We want to thank M. Morvan and the Laboratoire d' Informatique Algorithmique 
(LIAFA) of the University of Paris VII; this work was started during a visit 
of A. Moreira to that unit. We were also supported by the FONDAP
program in Applied Mathematics, and A. Moreira was partially supported by the CONICYT
Ph. D. fellowship.

\bibliographystyle{alpha}
\bibliography{ncca2}
\label{sec:biblio}

\end{document}